\documentclass[conference]{IEEEtran}
\usepackage{graphicx} 
\usepackage{algpseudocode}
\usepackage{algorithm}
\usepackage{amsmath, amssymb}
\usepackage{bm}
\usepackage{xfrac}
\usepackage{xspace}
\usepackage{hyperref}
\usepackage{makecell}
\usepackage[table]{xcolor}
\usepackage{subcaption}
\usepackage{booktabs}
\usepackage[most]{tcolorbox}


\newcommand{\rev}[1]{\textcolor{black}{#1}}

\newcommand{\kkm}{Kernel K-means\xspace}
\newcommand{\V}{\mathbf{V}\xspace}
\newcommand{\Pp}{\mathbf{P}\xspace}
\newcommand{\K}{\mathbf{K}\xspace}
\newcommand{\E}{\mathbf{E}\xspace}
\newcommand{\D}{\mathbf{D}\xspace}
\newcommand{\B}{\mathbf{B}\xspace}
\newcommand{\z}{\mathbf{z}\xspace}
\newcommand{\cv}{\mathbf{c}\xspace}
\newcommand{\R}{\mathbb{R}\xspace}
\newcommand{\allgather}{$\mathsf{Allgather}$\xspace}
\newcommand{\allreduce}{$\mathsf{Allreduce}$\xspace}
\newcommand{\alltoallv}{$\mathsf{Alltoallv}$\xspace}
\newcommand{\mpiallgather}{$\mathsf{MPI\_Allgather}$\xspace}
\newcommand{\mpibcast}{$\mathsf{MPI\_Bcast}$\xspace}
\newcommand{\mpiallreduce}{$\mathsf{MPI\_Allreduce}$\xspace}

\newcommand{\mpigather}{$\mathsf{MPI\_Gather}$\xspace}
\newcommand{\argmin}{$\mathsf{argmin}$\xspace}
\newcommand{\cO}{\mathcal{O}\xspace}

\makeatletter
\newcommand{\linebreakand}{%
  \end{@IEEEauthorhalign}
  \hfill\mbox{}\par
  \mbox{}\hfill\begin{@IEEEauthorhalign}
}
\makeatother

\title{Communication-Avoiding Linear Algebraic Kernel K-Means on GPUs}

\author{
\IEEEauthorblockN{Julian Bellavita}
\IEEEauthorblockA{Cornell University\\Ithaca, NY, USA\\jb2695@cornell.edu}
\and
\IEEEauthorblockN{Matthew Rubino}
\IEEEauthorblockA{Cornell University\\Ithaca, NY, USA\\mrr245@cornell.edu}
\and
\IEEEauthorblockN{Nakul Iyer}
\IEEEauthorblockA{Cornell University\\Ithaca, NY, USA\\npi2@cornell.edu}
\and
\IEEEauthorblockN{Andrew Chang}
\IEEEauthorblockA{Cornell University\\Ithaca, NY, USA\\yc723@cornell.edu}
\linebreakand 
\IEEEauthorblockN{Aditya Devarakonda}
\IEEEauthorblockA{Wake Forest University\\Winston-Salem, NC, USA\\devaraa@wfu.edu}
\and
\IEEEauthorblockN{Flavio Vella}
\IEEEauthorblockA{University of Trento\\Trento, Italy\\flavio.vella@unitn.it}
\and
\IEEEauthorblockN{Giulia Guidi}
\IEEEauthorblockA{Cornell University\\Ithaca, NY, USA\\gguidi@cornell.edu}
}

\begin{document}

\maketitle

\begin{abstract}
Clustering algorithms are among the most important and informative data analysis tools available.
Of these, the K-means algorithm is one of the most popular and powerful choices due to its simplicity and generality. However, K-means cannot compute non-linearly separable cluster boundaries, which limits its utility in certain cases. \kkm is a variant of the K-means algorithm that can compute non-linearly separable cluster boundaries. \kkm has significant computational and memory requirements because of the presence of a large kernel matrix which scales quadratically with dataset size. 
Prior work has accelerated \kkm by formulating it using sparse linear algebra primitives and implementing it on a single GPU. However, that approach cannot run on datasets with more than approximately 80,000 samples due to limited GPU memory.

Clustering using \kkm remains challenging because of its high computational and memory costs. In this work, we address this issue by presenting a suite of distributed-memory parallel algorithms for large-scale \kkm clustering on multi-GPU systems. Our approach maps the most computationally expensive components of \kkm onto communication-efficient distributed linear algebra primitives uniquely tailored for \kkm, enabling highly scalable implementations that efficiently cluster million-scale datasets.
\rev{ Central to our work is the design of partitioning schemes that enable communication-efficient composition of the linear algebra primitives that appear in \kkm.} 

\rev{Our 1.5D algorithm consistently achieves the highest performance, enabling \kkm to scale to data one to two orders of magnitude larger than previously practical.} 
On 256 GPUs, it achieves a geometric mean weak scaling efficiency of $79.7\%$ and a geometric mean strong scaling speedup of $4.2\times$.
Compared to our 1D algorithm, the 1.5D approach achieves up to a $3.6\times$ speedup on 256 GPUs and reduces clustering time from over an hour to under two seconds relative to a single-GPU sliding window implementation.
Our results show that distributed algorithms designed with application-specific linear algebraic formulations can achieve substantial performance improvement.
    
\end{abstract}

\section{Introduction}

K-means is a widely used unsupervised learning algorithm that attempts to group similar points to identify patterns in data. It is applied in biology~\cite{bio-clustering}, economics~\cite{econ-clustering}, and computer vision~\cite{cv-clustering}. However, it cannot capture non-linearly separable clusters, which can degrade clustering quality~\cite{dhillon2004kernel}.

\kkm is a K-means variant that can compute non-linearly separable clusters, often resulting in higher quality clustering for non-linearly separable data~\cite{ferrarottietal,gonen2014localized,zhang2002large}. 
It achieves this by clustering points in a high-dimensional feature space, producing non-linear boundaries in the original space. 
This is enabled by the kernel trick, which computes inner products between points in feature space using a non-linear kernel applied to the original points. 
The resulting values are stored in a kernel matrix, $\mathbf{K}$, avoiding the need to explicitly project points and thus saving computation and memory.

A drawback of \kkm is that its per-iteration time complexity increases quadratically with the number of points, making it costly even for medium-sized data. To address this, prior work demonstrates effective GPU parallelization by mapping the main clustering loop to sparse linear algebra and using dense linear algebra for kernel matrix computation~\cite{popcorn}.
The main limitation of this approach is that it cannot cluster datasets with more than approximately eighty thousand points due to limited GPU memory and the large size of the kernel matrix.
Datasets such as MNIST8m~\cite{libsvm} contain millions of points, and applications like molecular dynamics~\cite{ferrarottietal} and human activity recognition~\cite{jamel2020human} often require clustering at this scale. 
As datasets in scientific and engineering domains continue to grow~\cite{tosi202415}, single-GPU approaches to \kkm are insufficient in these big data scenarios.

\rev{
A common approach to scaling \kkm is low-rank approximation, such as the Nystr\"{o}m method, which does not explicitly form the kernel matrix. 
The approximate approaches are generally more scalable than the exact approach, but they can yield poor outcomes for certain kernel matrices, such as those with slow spectral decay or high coherence \cite{radhaetal, rebrova2018studyclusteringtechniqueshierarchical, chen2024randomlypivotedcholeskypractical}.
Moreover, they often require extensive dataset- and $k$-dependent tuning, sometimes relying on access to the full kernel matrix \cite{wang2019scalablekernelkmeansclustering}. In contrast, exact \kkm can be used universally without tuning.
}

Here, we tackle exact \kkm clustering on million-scale datasets through GPU acceleration and distributed memory, \rev{expanding the practical scale of \kkm by one to two orders of magnitude.}
Our approach relies on a mapping of \kkm to communication-efficient sparse and dense linear algebra primitives. Through this mapping, we introduce new distributed \kkm algorithms that reduce communication by explicitly exploiting domain-specific structure \rev{and effectively composing the distribution schemes of the different linear algebra primitives involved in \kkm.}

\rev{
The linear algebraic formulation of \kkm consists of a dense general matrix multiplication (GEMM) followed by multiple sparse-dense matrix multiplications (SpMM), where the dense matrix in the SpMM is the output of the GEMM.
There is a rich design space of distributed SpMM and GEMM algorithms to consider, each with distinct distribution strategies and communication schedules ~\cite{brock2024rdma,cosma,sparse-summa, summa}. 
In this work, we develop 1D, Hybrid 1D, 1.5D, and 2D distributed \kkm algorithms, each tailored to exploit domain-specific characteristics of \kkm.
Central to our approach is the composability, or fusion, of the GEMM and SpMM primitives. 
Rather than considering each operation in a vacuum, our most performant algorithm chooses partitioning schemes and communication schedules for both primitives that enable efficient communication when considered together.
In addition, the linear algebraic formulation of \kkm uses a sparse matrix $\V$ with exactly one nonzero per column, which we leverage to design partitioning schemes and communication schedules that achieve perfect load balance while minimizing communication.
}

Our 1D algorithm efficiently executes the clustering loop but cannot compute the kernel matrix at scale. 
It is described as a baseline because its communication pattern resembles prior distributed \kkm approaches~\cite{farivar2008parallel, tsapanos2015distributed} that do not use distributed sparse primitives and are not publicly available, so a direct comparison is not possible.
The Hybrid 1D algorithm addresses some scalability issues of the 1D algorithm, but it introduces significant additional memory and communication overhead, making it ineffective in some cases. 
The 2D algorithm is scalable and memory-efficient, but its 2D partitioning of the distance matrix introduces communication overhead when updating cluster assignments.

Our main contribution is the 1.5D algorithm, which communicates asymptotically less than other approaches and makes it feasible to use exact \kkm to cluster million-scale datasets. 
The 1.5D algorithm is different from existing approaches to distributed \kkm, which are largely similar to our 1D algorithm. 
\rev{A key novelty of the 1.5D algorithm is its choice of partitioning schemes for the input matrices, which enables efficient composition of GEMM and SpMM, allowing both to be computed in a communication-efficient manner. 
More broadly, our results highlight application-guided composition of distributed linear algebra primitives as a key research direction beyond individual scalable primitives.}

Our algorithms were evaluated on three datasets and compared to a single-GPU sliding-window implementation. The 1.5D and 2D algorithms scale to 256 GPUs. The 1.5D algorithm outperforms the 1D algorithm by up to $3.6\times$ in strong scaling and by over $2000\times$ compared to the sliding-window approach. 
Notably, both 1.5D and 2D can process over $1.5$ million points without exhausting memory.
Our algorithms are implemented in the open-source software \textsc{Vivaldi} (named after the composer) and are available at \url{https://github.com/CornellHPC/Vivaldi}.
The sliding window algorithm is available at \url{https://github.com/aditya08/sliding-window-kernel-kmeans}.

\noindent
In summary, our main contributions are:
\begin{enumerate}
    \item The first open-source fully GPU-accelerated distributed memory implementation of exact \kkm;
    \item A set of new parallel algorithms for \kkm based on distributed linear algebra primitives;
    \item \rev{Our 1.5D algorithm achieves up to $3.6\times$ speedup over the baseline and makes clustering of million-scale datasets practical by increasing the dataset size limit by one to two orders of magnitude.}
\end{enumerate}
\section{Background}\label{sec:background}

\subsection{\kkm Clustering}

K-means is an unsupervised learning algorithm that partitions data into $k$ clusters by assigning each point to the nearest centroid using a chosen distance metric, typically Euclidean~\cite{lloydskmeans}. \kkm extends K-means to non-linearly separable data by performing clustering in a higher-dimensional feature space and mapping the resulting centroids back to the input space.
\kkm updates centroids iteratively by (1) computing point–centroid distances in feature space, (2) assigning each point to its nearest centroid, and (3) recomputing centroids as the mean of the assigned points.
In practice, feature-space projections are never computed explicitly. 
Instead, \kkm uses a kernel function $\kappa(x, y)$ to evaluate inner products in feature space while operating in the input space.
Pairwise distances are derived from $\kappa(x, y)$, resulting in $\cO(n^2)$ work per iteration, where $n$ is the number of points in the dataset~\cite{dhillon2004kernel}.

\subsection{\kkm with Sparse Linear Algebra}

The state-of-the-art method for parallelizing \kkm on a GPU formulates the problem using GEMM and the sparse linear algebra primitives SpMM and sparse matrix-vector multiplication (SpMV). The components of this formulation relevant to this work are summarized below; further details are available in \cite{popcorn}.

Let $\Pp \in \R^{n \times d}$ be the (dense) point matrix, where $n$ is the number of points and each row of $d$ features represents one point.
Then, we define the symmetric matrix $\B \in \R^{n \times n}$ as:
\begin{equation}\label{eq:b}
    \B = \Pp\Pp^T 
\end{equation}

\noindent
The kernel matrix $\K \in \R^{n \times n}$ is obtained by applying an element-wise kernel function to $\B$. 
For example, if we define our kernel function $\kappa$ to be the polynomial kernel:
\begin{equation}\label{eq:k}
    \kappa(x, y) = (\gamma x^Ty + c)^d 
\end{equation}
where $x, y \in \R^d$, then we have $\K(i,j) = \kappa(\Pp(i, :), \Pp(j, :))$, which is equivalent to $\K(i,j) = (\gamma\B(i,j) + c)^d$, therefore
$\K$ can be obtained by applying $\kappa$ to each entry of $\B$. 
In practice, $\K$ is usually large, and the cost of storing or computing with $\K$ is the main bottleneck in \kkm.

Once $\K$ is computed, the clustering loop can begin. 
Let $k$ be the number of clusters, initialized by some strategy.
The way we initialize the clusters does not affect the clustering loop computation; it only affects convergence properties \cite{kmeanspp}.
Define the sparse matrix $\V \in \R^{k \times n}$:
\rev{\begin{equation}\label{eq:v}
    \V(i,j) = \begin{cases}
        \frac{1}{|L_i|} & \text{if point $j$ is in cluster $i$} \\
        0 & \text{otherwise}
    \end{cases}
\end{equation}
}

\noindent
Let $L_i$ denote the points in cluster $i$. Notably:
\begin{enumerate}
    \item $\V$ is a sparse matrix with exactly $n$ nonzeros;
    \item $\V$ has exactly 1 nonzero per column.
\end{enumerate}

\noindent
First, each clustering iteration computes:

\begin{equation}\label{eq:e}
   \E = \K\V^T 
\end{equation}

\noindent
It then initializes $\mathbf{z} \in \mathbb{R}^n$ using a masking operation on $\mathbf{E}$:

\begin{equation}\label{eq:z}
    \mathbf{z}(i) = \E(i, cl(i)) 
\end{equation}

\noindent
$cl(i)$ is a function that returns the cluster to which point $i$ is assigned.
Then, a matrix-vector product is computed:

\begin{equation}\label{eq:c}
    \mathbf{c} = \V\z
\end{equation}

\noindent
\rev{The vector $\mathbf{c}$ is equal to a single row of the matrix $\tilde{\mathbf{C}} \in \mathbb{R}^{n \times k}$, defined as:
\begin{equation}\label{eq:ct}
    \tilde{\mathbf{C}} = \begin{bmatrix}
        \|c_1\|_2^2 & \|c_2\|_2^2 & \dots & \|c_k\|_2^2 \\
        \|c_1\|_2^2 & \|c_2\|_2^2 & \dots & \|c_k\|_2^2 \\
        \vdots & \vdots & \vdots & \vdots \\
        \|c_1\|_2^2 & \|c_2\|_2^2 & \dots & \|c_k\|_2^2 
    \end{bmatrix} 
\end{equation}
Given that each row of $\tilde{\mathbf{C}}$ is identical, $\mathbf{c}$ can serve as a compact representation of $\tilde{\mathbf{C}}$.
}
The distance matrix $\D \in \R^{n\times k}$ is then computed as:

\begin{equation}\label{eq:d}
    \D = -2\E + \mathbf{\tilde{C}} 
\end{equation}

\noindent
$\D(i,j)$ stores the distance between point $i$ and cluster $j$ in feature space.  
Cluster assignments are updated via a row-wise argmin and used to update $\V$, after which the next iteration can begin.


In practice, \eqref{eq:b} is computed using GEMM, \eqref{eq:e} using SpMM, and \eqref{eq:c} using SpMV. Thus, the main parts of \kkm reduce to linear algebra primitives with efficient parallel implementations in libraries such as cuSPARSE and cuBLAS \cite{cublas,cusparse}. 
Extending \kkm to distributed memory is a non-trivial task. The main challenges are avoiding communication of $\K$ partitions, which require $\mathcal{O}(n^2)$ bytes, and distributing $\Pp, \K, \V, \E$ to perform GEMM, SpMM, and cluster updates without excessive communication or load imbalance.

\subsection{Distributed Linear Algebra Primitives}

Prior work has developed parallel algorithms for linear algebra primitives on distributed memory.
For GEMM, a widely used approach is the Scalable Universal Matrix Multiply Algorithm (SUMMA) \cite{summa}, which partitions the input matrices $\mathbf{A}$ and $\mathbf{B}$ over a 2D process grid and performs GEMM as a sequence of distributed block outer products.
If $P$ is the total number of processes, SUMMA requires $\sqrt{P}$ rounds of communication and local computation. 
At iteration $p$, the process with row rank $p$ broadcasts its tile of $\mathbf{A}$ along its process row, while the process with column rank $p$ broadcasts its tile of $\mathbf{B}$ along its process column. 
Each process then multiplies the received tiles and accumulates the result into its local tile of $\mathbf{C}$. 
If we assume a tree-based broadcast, the communication cost for square matrices of dimension $n$ under the $\alpha$-$\beta$ model~\cite{hockney94} is:
\begin{equation}
    T_{SUMMA} = \alpha\cO(\sqrt{P}\log(\sqrt{P})) + \beta\cO(\log(\sqrt{P})\frac{n^2}{\sqrt{P}})
\end{equation}

\noindent
This is within a $\log(\sqrt{P})$ factor of the standard communication lower bound for dense linear algebra, assuming no additional memory is used for replication \cite{ballard2011minimizing}.

For SpMM, two widely used algorithms are 1.5D SpMM and 2D SpMM \cite{sparse-summa}. 
2D SpMM is algorithmically identical to SUMMA, with the only difference being that the first operand $\mathbf{A}$ is sparse. 
The different communication schedules for 2D SpMM are called A-stationary, B-stationary, and C-stationary. These schedules differ based on which matrices are communicated: A-stationary algorithms communicate only partitions of $\mathbf{B}$ and $\mathbf{C}$; C-stationary communicates only $\mathbf{B}$ and $\mathbf{A}$, and so on.
1.5D SpMM is a communication-avoiding algorithm that replicates partitions of one operand across multiple processes, reducing communication costs at the expense of higher per-process memory requirement. 
In the literature, only A-stationary variants of 1.5D SpMM that replicate partitions of $\mathbf{B}$ are described \cite{sparse-summa}. 

For GEMM and SpMM, there are 3D algorithms that divide the matrices on a 3D process cube~\cite{tripathy2020reducingcommunicationgraphneural, agarwal1995three} and communicate asymptotically less than 2D or 1.5D algorithms. 
3D algorithms are not considered in this paper for the following reasons. 
Using 3D GEMM to compute $\K$ would require replicating tiles of $\K$, further increasing its memory footprint.
Similarly, 3D SpMM is unlikely to improve the computation of $\E$, as it generally performs worse than 2D SpMM unless a very large number of processes is used~\cite{tripathy2020reducingcommunicationgraphneural}.


\section{Related Work}

The standard K-means algorithm has been used for several decades~\cite{lloydskmeans}, and prior work has studied its implementation on various parallel architectures~\cite{balcanetal,farivar2008parallel}. To the best of our knowledge, \kkm was first introduced by Girolami~\cite{girolami2002mercer}. Previous work on high-performance \kkm has explored parallelization in shared memory~\cite{baydounetal, popcorn} and distributed memory~\cite{ferrarottietal, tsapanos2015distributed}, but has not leveraged distributed linear algebra primitives. To our knowledge, these prior implementations are not open-source, and none are fully GPU-accelerated, making direct comparison difficult. \rev{The communication schedules of these approaches closely resemble those of our 1D algorithm, which is used as a practical baseline.} 
In Section~\ref{sec:results}, we show that the 1D algorithm is almost universally outperformed by our other approaches.

\rev{To scale \kkm, previous approaches stored the kernel matrix on disk~\cite{zhang2002large}, while low-rank approximations such as Nystr\"{o}m reduced memory usage~\cite{radhaetal}. 
However, these approximations can degrade clustering quality for kernel matrices with slow spectral decay or high coherence~\cite{chen2024randomlypivotedcholeskypractical,radhaetal, rebrova2018studyclusteringtechniqueshierarchical}.
They also introduce dataset- and $k$-dependent tuning parameters that are difficult to optimize at scale. 
FedKKM~\cite{zhou2022memory} constructed low-dimensional random feature approximations on edge devices using communication-efficient Lanczos algorithms, reducing memory usage by up to 94\% but increasing the number of iterations and computational overhead, with clustering quality dependent on approximation accuracy. 
Choosing appropriate approximations and tuning remains challenging, whereas exact \kkm eliminates both issues by computing the solution directly, making it universally effective.}
Prior GPU-based approaches~\cite{baydounetal, popcorn} are limited to single-device memory, whereas our work scales exact \kkm to million-point datasets, delivering high-quality clustering without tuning.

\rev{In addition, algorithms such as spectral clustering~\cite{dhillon2004kernel}, DBSCAN~\cite{gholizadeh2021k,deng2020dbscan}, and k-NN clustering~\cite{lulli2015scalable} can also capture nonlinear cluster structures, with k-NN graph-based clustering being particularly scalable by avoiding explicit kernel matrices. 
\kkm is a foundational algorithm with a strong theoretical basis and widespread practical use~\cite{handhayani2015intelligent,wang2019scalablekernelkmeansclustering, zhang2018does}, which is why it is the focus of this work.}

Prior work has explored distributed-memory algorithms that use linear algebra, particularly sparse primitives, including graph algorithms~\cite{azad2019lacc, azad2015parallel, combblas2, combblas, kepner2011graph, kimmerer2024graphblas,solomonik2017scaling}, Markov clustering~\cite{azad2018hipmcl}, GNN training~\cite{tripathy2020reducingcommunicationgraphneural, tripathy2024distributed}, and computational biology pipelines~\cite{besta2020communication, guidi2022distributed, guidi2021parallel, selvitopi2022extreme}.
\rev{
The composition and optimization of sequences of linear algebra primitives has recently emerged as a key research area in high-performance computing. 
Bharadwaj et al.~\cite{bharadwaj2022} develop communication-avoiding strategies for sequences of Sampled Dense-Dense Matrix Multiply (SDDMM) and SpMM operations in GNNs, showing that, in that application, performance depends on how communication schedules are composed across consecutive primitives. 
Compiler-based BLAS fusion~\cite{filipovivc2015optimizing} and DNN operator fusion~\cite{niu2021dnnfusion} similarly demonstrate that optimizing primitives in isolation can leave substantial performance unrealized.
In \kkm, we show that adapting GEMM and SpMM sequences to sparsity and primitive composition reduces communication and improves end-to-end performance, highlighting composability as a key research direction in high-performance linear algebra.}


\section{Algorithm Descriptions}\label{sec:alg}


Here, we describe our distributed-memory \kkm algorithms. 
Let $P$ be the total number of processes. 
For simplicity, assume that $P$ evenly divides $n$, $\sqrt{P}$ evenly divides $k$, and that 2D process grids are $\sqrt{P} \times \sqrt{P}$, although only the square process grid assumption is required for correctness. 
In a 2D process grid, the $i$th process in column $j$ is denoted $P(i, j)$. 
$\K_{ij}$ is the portion of $\K$ on $P(i, j)$, and $\cv_i$ is the portion of $\cv$ on the process with global rank $i$. 
For simplicity, we set $\B = \K$, corresponding to the linear kernel (i.e., standard dot product); this simplifies notation without affecting the algorithms. 
Communication costs are analyzed using the standard $\alpha$-$\beta$ model~\cite{hockney94}, where $\alpha$ represents message latency and $\beta$ denotes network bandwidth.

Each algorithm is defined by its strategy for partitioning $\Pp$, $\K$, $\V$, and $\E$, as well as its approach to distributed GEMM and SpMM. Our focus is on performing the following operations in distributed memory:
\begin{enumerate}
 \item {Distributed GEMM}: $\K = \Pp\Pp^T$
 \item {Distributed SpMM}: $\E^T = \V\K$
\end{enumerate}

$\E^T$ is computed instead of $\E$ because SpMM libraries typically require the first operand to be sparse; this does not affect the algorithm. 
GEMM multiplies a tall, narrow matrix by a short, wide matrix, while SpMM multiplies a sparse matrix with one nonzero per column by a large, square dense matrix.


\rev{
\kkm has application-specific characteristics that must be considered for efficient distributed-memory parallelization, including the unique sparsity structure of $\mathbf{V}$, the comparatively large size of $\mathbf{K}$, and the need to perform cluster updates at each iteration. 
To address these challenges, we present four algorithms that use different distribution strategies based on linear algebra.}

\rev{
In our 1D algorithm, all matrices are partitioned into 1D column blocks. This results in an algorithm that can compute the SpMM in a load-balanced manner without communicating $\K$, but the 1D layout of $\K$ leads to high communication costs for both GEMM and SpMM.
Two natural alternatives reduce GEMM communication: a Hybrid 1D algorithm that redistributes $\K$ from 2D to 1D, and a 2D algorithm that modifies the partitioning of $\V$ to be 2D.
Both algorithms use SUMMA to compute $\K$, which reduces the communication costs of the GEMM.
However, the Hybrid 1D algorithm incurs a high redistribution cost from 2D to 1D, while the 2D algorithm requires additional communication when updating cluster assignments.
}

\rev{
The need to perform both a distributed GEMM and a distributed SpMM sequence motivates our main contribution: a new 1.5D algorithm.
By using SUMMA for GEMM, the 1.5D algorithm preserves the natural 2D output partitioning of $\K$ and combines it with a 1D partitioning of $\mathbf{V}$, enabling a communication-efficient algorithm for distributed SpMM without redistribution and allowing cluster updates to be computed without additional communication.
This design gives 1.5D the lowest communication cost among the approaches considered.}

\begin{table*}[h!]
\centering
\begin{tabular}{lcccc}
\toprule
 & \textbf{1D} & \textbf{Hybrid 1D} & \textbf{1.5D} & \textbf{2D}\\
\midrule
Kernel Matrix ($\K$) & 
$\alpha\cO(P) + \beta\cO(Pnd)$  & 
$\alpha\cO(P) + \beta\cO(\frac{n^2}{P} + \frac{nd}{\sqrt{P}})$ &
$\alpha\cO(\sqrt{P}) + \beta\cO(\frac{nd}{\sqrt{P}})$ & 
Same as 1.5D \\
Distances Matrix ($\D^T$) & 
$\alpha\cO(P) + \beta\cO(n)$ & 
Same as 1D & 
$\alpha\cO(\sqrt{P}) + \beta\cO(\frac{n(k+1)}{\sqrt{P}})$ &
$\alpha\cO(\sqrt{P}) + \beta\cO(\frac{n(k+1)}{\sqrt{P}} + n)$\\
\bottomrule
\end{tabular}
\caption{Communication cost of $\K$ and $\D^T$ computation for each algorithm. $\log(\sqrt{P})$ terms are omitted for brevity.
}
\label{tab:comm}
    \vspace{-1em}

\end{table*}

\subsection{1D Kernel K-means}

Our first algorithm divides matrices in a 1D columnwise manner. Each process owns a contiguous block of $\frac{n}{P}$ columns from each matrix. $\Pp^T$ is initially partitioned as:
\begin{equation}
    \Pp^T = \begin{bmatrix}
       \Pp^T_{1} & \Pp^T_2 & \dots & \Pp^T_P
    \end{bmatrix}
\end{equation}

\noindent
Therefore, a given tile $\Pp_i^T \in \R^{\frac{n}{P} \times d}$ is stored on process $i$.
The first step of the algorithm computes $\K = \Pp \Pp^T$ using a 1D distributed GEMM. 
In this scheme, an \allgather collective replicates $\Pp$ on each process. 
Each process then computes a local GEMM between its local partition of $\Pp^T$ and the replicated $\Pp$, 
producing a block column of the kernel matrix $\K$.
$\K$ is partitioned columnwise as:

\begin{equation}
    \K = \begin{bmatrix}
        \K_1 & \K_2 & \dots & \K_P
    \end{bmatrix}
\end{equation}

\noindent
The main clustering loop then begins, with $\V$ initially partitioned columnwise as:
\begin{equation}
    \V = \begin{bmatrix}
        \V_1 & \V_2 & \dots & \V_P
    \end{bmatrix}
\end{equation}

\noindent
Each partition of $\V$ has exactly $\frac{n}{P}$ nonzeros, making communication inexpensive compared to the $\frac{n^2}{P}$ entries in each partition of $\K$. 
To compute $\E^T$, we use a 1D B-stationary distributed SpMM that communicates only $\V$. 
This requires a single \allgather to replicate $\V$ across processes, followed by a local SpMM on each process to compute its corresponding partition of $\E^T$. 
As a result, $\E^T$ is partitioned as:

\begin{equation}
    \E^T = \begin{bmatrix}
        \E^T_1 & \E^T_2 & \dots & \E^T_P
    \end{bmatrix}
\end{equation}

\noindent
Each process then performs the masking operation to create $\z_p$ locally, followed by a local SpMV and a global \allreduce to obtain $\cv$ on every process.
Finally, each process computes its partition of the distance matrix $\D^T$ and updates its cluster assignments by performing a local \argmin over the columns of $\D^T_p$.
The 1D algorithm is summarized in Algorithm~\ref{alg:1d}.

\begin{algorithm}[t]
    \caption{1D \kkm Algorithm}
    \label{alg:1d}
    \begin{algorithmic}[1]
        \Require $\Pp \in \R^{n \times d}$ stores points; number of clusters $k$. 
        \State \allgather $\Pp$ on each process \Comment{1D GEMM}
        \State All processes $p$ compute $\K_p = \Pp^T_p\Pp$
        \While{not converged} \Comment{Clustering loop}
            \State \allgather $\V$ on each process
            \State All processes $p$ compute $\E_p^T = \V\K_p^T$ 
            \State $\z_p = \text{mask}(\E^T_p)$
            \State $\cv_p = \V_p\z_p$
            \State \allreduce $\cv_p$ to compute $\cv$ on each process
            \State $\D^T_p = \text{distances}(\E^T_p, \cv)$
            \State $cl_p = \text{\argmin} (\D^T_p)$ \Comment{Updated cluster assignment}
            \State Update $\V_p$ using $cl_p$
        \EndWhile
    \end{algorithmic}
\end{algorithm}

The \allgather in the 1D GEMM routine involves $\mathcal{O}(P)$ messages and sends a total of $\mathcal{O}(Pnd)$ words. 
Thus, the communication cost of computing $\K$ is:

\begin{equation}
    T_{K1D} = \alpha \cO(P) + \beta \cO(Pnd)
\end{equation}

\noindent
The \allgather used to compute $\E^T$ replicates $\V$ on each process. 
If we assume a pairwise exchange algorithm is used for the \allgather~\cite{thakur2005optimization}, this involves a total of $\cO (P)$ messages. Since $\V$ is sparse with exactly $n$ nonzeros, the total number of words communicated is $\mathcal{O}(n)$, resulting in a total communication cost for computing $\E^T$ of:

\begin{equation}
    T_{E1D} = \alpha\cO(P) + \beta\cO(n)
\end{equation}

\noindent
The only other communication is the global \allreduce for $\cv$, a vector of length $k$, which is negligible since $k$ is small.

The main benefit of the 1D algorithm arises in the clustering loop, where columnwise partitioning of $\V$ and $\K$ provides perfectly load-balanced SpMM due to the uniform nonzeros in the replicated $\V$.
Perfect load balance is typically difficult to achieve in distributed sparse linear algebra~\cite{ brock2024rdma,hong2024sparsityawaredistributedmemoryalgorithmsparsesparse, sparse-summa}, making this a notable advantage of the 1D algorithm. 
In addition, updating cluster assignments requires no communication.

Overall, the communication cost of the 1D algorithm does not scale with the number of processes. 
The 1D GEMM requires an \allgather of $\mathcal{O}(Pnd)$ words, which becomes a bottleneck for large $P$.
\rev{Similarly, the \allgather for $\E^T$ sends $\mathcal{O}(n)$ words; although this amount does not increase as $P$ increases, it may dominate when local computation becomes negligible or in a weak scaling setting.}
The 1D algorithm also has a large memory footprint: replicating $\Pp$ on each process can cause out-of-memory errors when $d$ is large, especially because partitions of $\K$ must also be stored.
\rev{The limitations of the 1D algorithm arise from the communication costs of both GEMM and SpMM.}

\begin{figure*}[t]
    \centering
    \includegraphics[width=\linewidth]{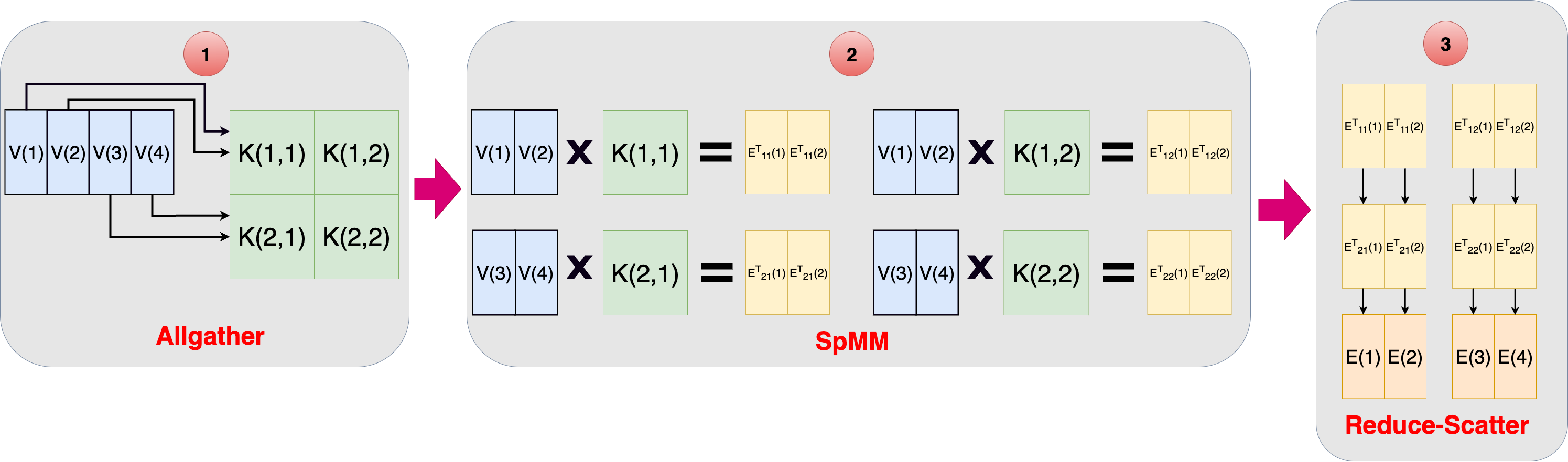}
    \caption{1.5D SpMM algorithm on $P=4$ processes. \rev{$\V$ is partitioned 1D columnwise and $\K$ in 2D. (1) The nonzeros of each $\V$ partition are replicated along the corresponding process row. (2) Each process performs a local SpMM with its $\V$ replicas and local $\K$ partition, producing partial sums of $\E^T$. (3) The partial sums are split along columns and reduced across process columns, resulting in a 1D columnwise partitioning of $\E^T$.}}
    \label{fig:15d}
        \vspace{-1em}
\end{figure*}


\subsection{Two Alternatives: Hybrid 1D and 2D}
\rev{
To reduce GEMM communication in the 1D algorithm, the 1D GEMM can be replaced with SUMMA~\cite{summa}, which computes distributed GEMM with a near-optimal communication cost.
However, using SUMMA results in $\K$ being partitioned in a 2D distribution. This prevents the use of the approach from the 1D algorithm during the main clustering loop, since $\K$ is no longer distributed column-wise in 1D. Therefore, changes to the algorithm for computing the SpMM and cluster updates in the main clustering loop are necessary. Here, we consider two algorithms that implement these changes: a Hybrid 1D algorithm (H-1D) and a pure 2D algorithm.
}

\rev{In H-1D, SUMMA is used to compute $\K$. 
Then, $\K$ is redistributed from SUMMA's 2D layout to a 1D column-wise distribution, after which the main clustering loop is identical to that of the 1D algorithm.
This redistribution is performed using an \alltoallv.
Computing $\K$ using SUMMA incurs the following communication cost:
\begin{equation}
    T_{K-SUMMA} = \alpha\cO\bigl(\sqrt{P}\log\bigl(\sqrt{P}\bigr)\bigr) + \beta\cO\biggl(\log\bigl(\sqrt{P}\bigr)\frac{nd}{\sqrt{P}}\biggr)
\end{equation}
H-1D enables a more communication-efficient algorithm for performing GEMM, but it incurs additional communication costs due to the redistribution of $\K$.
In particular, during the redistribution step, each process sends $\cO(P)$ messages and communicates at most $\cO(n^2 / P^{3/2})$ words to up to $\sqrt{P}$ processes, resulting in a worst-case redistribution cost of:
\begin{equation}
    T_{Redist} = \alpha\cO(P) + \beta\cO\biggl(\frac{n^2}{P}\biggr)
\end{equation}
This $\cO(n^2 / P)$ redistribution volume can make the H-1D algorithm communication costly unless a very large number of processes are used. Furthermore, the scalability issues of SpMM in the 1D algorithm remain present in H-1D.}

\rev{The second alternative we consider is a pure 2D approach, where both $\V$ and $\K$ are partitioned across a 2D process grid. 
This enables the use of SUMMA to compute $\K$ without needing to redistribute $\K$ from a 1D to a 2D layout before starting the clustering loop.
The computation of $\E^T$ uses a B-stationary SpMM~\cite{sparse-summa}, communicating only $\V$ entries and partial sums of $\E^T$. 
This results in a communication cost of:
\begin{equation}
    T_{E2D} = \alpha\cO\bigl(\sqrt{P}\bigr) + \beta\cO\biggl(\frac{n(k+1)}{\sqrt{P}}\biggr)
\end{equation}
The communication for computing $\E^T$ now scales with the processes; however, the 2D partitioning of $\E^T$ introduces extra communication for cluster updates. 
In particular, computing each point's nearest centroid requires an \argmin reduction along process columns, and an \allreduce along process rows to compute cluster sizes.
This results in a total cluster update communication of:
\begin{equation}
    T_{update} = \alpha\cO\bigl(\log(\sqrt{P})\bigr) + \beta\cO\biggl(\log(\sqrt{P})n\biggr)
\end{equation}
\noindent
}
\rev{The pure 2D algorithm has an asymptotically smaller communication volume than either 1D algorithm, but it requires communication during cluster updates, resulting in an overall communication cost higher than that of the 1.5D algorithm.
}


\subsection{1.5D Kernel K-means}

\rev{
Here, we present our 1.5D distributed \kkm algorithm, which can compute both the GEMM and SpMM in a scalable fashion.
The key idea behind our 1.5D algorithm is to use SUMMA to compute $\K$ while keeping $\V$ in a 1D distribution.
The subsequent distributed SpMM, $\E^T = \V\K$, is performed with $\V$ 1D-distributed and $\K$ 2D-distributed, enabling communication costs that decrease with process count and avoiding redistribution of $\K$ to 1D. 
The SpMM is further structured to produce $\E^T$ in a 1D columnwise distribution, allowing cluster updates to proceed without additional communication. 
This approach achieves scalable GEMM and SpMM while avoiding the limitations of the other three algorithms.
}


For simplicity, the 1.5D approach is described using $P = 4$, although Algorithm~\ref{eq:15d} describes the general case.
\rev{
To avoid communicating large partitions of $\K$ during SpMM, we use a B-stationary communication schedule that exchanges only entries of $\V$ and partial sums of the output $\E^T$. 
The distributed SpMM begins with an \allgather that replicates subsets of the 1D-partitioned $\V$ along each process row. 
In particular, the first two partitions of $\V$ are replicated along the first process row, and the last two along the second process row: $\begin{bmatrix} \V_{1} & \V_{2} \end{bmatrix}$ are replicated on processes $P(1,1)$ and $P(1,2)$, while $\begin{bmatrix} \V_{3} & \V_{4} \end{bmatrix}$ are replicated on processes $P(2,1)$ and $P(2,2)$.
Each process then performs a local SpMM:
\begin{equation}
 \E_{ij}^T = \K_{ij} \begin{bmatrix}
 \V_{2(i-1)} & \V_{2(i-1) + 1}
 \end{bmatrix}
\end{equation}
Each partition of $\V$ contains $\frac{n}{P}$ nonzeros, and the same number of $\V$ partitions are replicated on each process. 
Thus, the local SpMMs are load balanced, as in the 1D algorithm.
}

\rev{
Each $\E^T_{ij}$ has $k$ rows and $\frac{n}{\sqrt{P}}$ columns and represents a partial sum of the output. To form the complete $\E^T$, partial sums $\E^T_{ij}$ associated with the same process column are accumulated. For example, when $P=4$, $\E^T_{11} + \E^T_{21}$ produces the first $\frac{n}{2}$ columns of $\E^T$, while $\E^T_{12} + \E^T_{22}$ produces the remaining $\frac{n}{2}$ columns. The resulting sums can then be partitioned along either rows or columns, which determines the final layout of $\E^T$. Prior 1.5D SpMM approaches~\cite{sparse-summa} partition along rows via a $\mathsf{Reduce\text{-}Scatter}$ across process columns, resulting in the following partitioning of each $\E^T_{ij}$:
\begin{equation}
    \E^T_{ij} = \begin{bmatrix}
        \E_{ij}^T(1) \\
        \E_{ij}^T(2) 
    \end{bmatrix}
\end{equation}
Partitions labeled $(1)$ are reduced on the first process in column $j$, and those labeled $(2)$ on the second. 
This results in a 2D partitioning of $\E^T$, which is undesirable because it increases communication for both cluster updates and $\V$ updates. 
A 1D columnwise partitioning of $\E^T$ would be preferable, as it would avoid this additional communication.
To achieve this, we modify the $\mathsf{Reduce\text{-}Scatter}$ operation so that each $\E^T_{ij}$ is partitioned along columns instead of rows:
\begin{equation}
    \E^T_{ij} = \begin{bmatrix}
        \E_{ij}^T(1) & \E_{ij}^T(2) 
    \end{bmatrix}
\end{equation}
Each $\E_{ij}^T(l)$ has $k$ rows and $\frac{n}{p}$ columns. 
A $\mathsf{Reduce\text{-}Scatter}$ is performed so that process $p$ reduces $\E^T_{ij}(l)$ with $2(j-1) + l = p$. 
$\E^T$ is fully computed and naturally 1D columnwise partitioned, which enables cluster updates and $\V$ updates to be computed without any communication, as in the 1D algorithm. 
Our 1.5D SpMM algorithm is shown in Figure \ref{fig:15d}.
}

\begin{algorithm}[t]
    \rev{
    \caption{1.5D \kkm Algorithm}
    \label{eq:15d}
    \begin{algorithmic}[1]
        \Require $\Pp \in \R^{n \times d}$ stores points; number of clusters $k$. 
        \State{Use SUMMA to compute $\K$, partitioned across 2D grid} 
        \While{not converged} \Comment{Clustering loop}
            \State{\allgather $\begin{bmatrix}
                \V_{(i-1)\sqrt{P}} & \dots & \V_{(i-1)\sqrt{P}+\sqrt{P}}
            \end{bmatrix}$ on $P(i, :)$, store in $\V_{gather}$}
            \State{Compute $\E^T_{ij} = \V_{gather}K_{ij}$}
            \State{Partition $\E_{ij}^T$ into $\begin{bmatrix}
                \E_{ij}^T(1) & \dots & \E_{ij}^T(\sqrt{P})
            \end{bmatrix}$}
            \State{Reduce-scatter along process columns. $\E^T_{ij}(l)$ is reduced on process with global rank $p=l+(j-1)\sqrt{P}$}
            \State{$\E^T$ is now partitioned along columns in a 1D fashion.}
            \State $\z_p = \text{mask}(\E^T_p)$
            \State $\cv_p = \V_p\z_p$
            \State \allreduce $\cv_p$ to compute $\cv$ on each process
            \State $\D^T_p = \text{distances}(\E^T_p, \cv)$
            \State $cl_p = \text{\argmin} (\D^T_p)$ \Comment{Updated cluster assignment}
            \State Update $\V_p$ using $cl_p$
        \EndWhile
    \end{algorithmic}
    }
\end{algorithm}

In the 1.5D algorithm, computing $\K$ has the same communication cost as H-1D, as it relies on SUMMA. 
The only difference in communication arises when computing $\E^T$. The \allgather that replicates $\V$ partitions along process rows, sending $\cO(\sqrt{P})$ messages and $\cO(n/P)$ words to $\sqrt{P}$ processes, resulting in a total communication cost of:

\begin{equation}
    T_{allgather} = \alpha\cO\bigl(\sqrt{P}\bigr) + \beta\cO\biggl(\frac{n}{\sqrt{P}}\biggr)
\end{equation}

\noindent
The bandwidth scales with the number of processes, unlike $T_{E1D}$ from the 1D algorithm, while the latency term is reduced by a factor of $\sqrt{P}$. The $\mathsf{Reduce\text{-}Scatter}$ used to finalize $\E^T$ sends $\cO(\log(\sqrt{P}))$ messages and communicates $\cO(\frac{nk}{P})$ words to $\sqrt{P}$ processes, yielding a total communication cost of:
\begin{equation}
    T_{reduce-scatter} = \alpha\cO\bigl(\log\bigl(\sqrt{P}\bigr)\bigr) + \beta\cO\biggl(\frac{nk}{\sqrt{P}}\biggr)
\end{equation}

\noindent
The overall communication cost of computing $\E^T$ is:

\begin{equation}
    T_{E1.5D} = \alpha\cO\bigl(\sqrt{P}\bigr) + \beta\cO\biggl(\frac{n(k+1)}{\sqrt{P}}\biggr)
\end{equation}

\noindent
Our 1.5D algorithm reduces communication compared to 2D because cluster assignment does not require any communication. The $\frac{nk}{\sqrt{P}}$ bandwidth term is larger than that in 1D approaches for small $P$, but it scales with the process count: for large $P$, it is less than the $\cO(n)$ bandwidth term for 1D. The 1.5D approach uses only $\cO(\sqrt{P})$ messages, resulting in lower latency costs than 1D, and leverages SUMMA for $\K$ without the redistribution overhead of H-1D, further reducing communication.

\rev{
The 1.5D algorithm is the most effective and demonstrates the importance of considering the composability of primitives in distributed algorithms that use linear algebra. 
By design, keeping $\V$ 1D-partitioned enables SpMM to efficiently consume the 2D-partitioned $\K$ produced by SUMMA without redistribution.
Moreover, keeping $\mathbf{E}^T$ 1D-partitioned eliminates communication for cluster updates, highlighting the importance of application-aware partitioning schemes.
}

\rev{
Our approach, developed for distributed \kkm, could also benefit other clustering algorithms that use GEMM and SpMM, such as K-means \cite{cuda-kmeans} and spectral clustering \cite{dhillon2004kernel}. 
In distributed memory, these algorithms could similarly benefit from communication schedules that emphasize composability.
}

\rev{
}


\section{Implementation}

This section describes the implementation of each proposed algorithm. The algorithms are written in C++, using dense matrices in row-major order and local $\V$ partitions in compressed sparse column (CSC) format. Storing dense matrices in row major order is known to improve the performance of cuSPARSE's SpMM routine \cite{cusparse}.
The implementation is GPU-capable, storing matrices and vectors on the device and using single-precision floating-point numbers and 32-bit integers for indices.
The local sparse operations use cuSPARSE \cite{cusparse}, distributed GEMM uses SLATE \cite{slate_doc}, and communication is managed through GPU-aware MPI.

For the 1D and 1.5D algorithms, communication of $\V$ partitions involves only their local row indices.
This is sufficient to initialize the partition of $\V$ on the receiving process, since the column pointers array contains the columns in the partition, and values can be computed using the global cluster sizes from an \allreduce at the end of each iteration.

In the 2D algorithm, the column pointers array is sent along with the local row indices, but values are initialized locally as in the 1D and 1.5D algorithms. 
$\V$ is initialized by assigning points to clusters in a round-robin fashion. 
Other initialization strategies that provide stronger convergence guarantees, such as K-Means++~\cite{kmeanspp}, are left for future work.

\subsection{1D and Hybrid 1D Implementation}

The 1D implementation begins by initializing $\Pp^T$ and $\Pp$ on the devices. In the pure 1D implementation, both matrices are partitioned in a 1D fashion, while in the H-1D implementation, both matrices are partitioned in a 2D fashion. 
From here, SLATE's GEMM routine computes $\K$. For the H-1D algorithm only, $\K$ is then redistributed to a 1D columnwise partitioning using \alltoallv. From this point, the 1D and H-1D implementations proceed identically. $\V$ is initialized using the round-robin assignment strategy.

$\V$ is communicated via \mpiallgather, followed by partitioned $\E^T$ computation using cuSPARSE SpMM. 
The masking for $\z$ uses a custom kernel, and $\cv$ is computed with cuSPARSE SpMV and finalized with a global \mpiallreduce.
A hand-written kernel sums $\tilde{\mathbf{C}}$ using $\cv$ and $\E^T$ to form $\D^T$. Then, local \argmin values are computed with another kernel, producing partitions of the cluster assignments array, which also serve as row indices for local $\V$ partitions. Finally, a global \allreduce computes cluster sizes for updating $\V$.

\subsection{2D Implementation}

In our 2D approach, $\Pp^T$, $\Pp$, and $\K$ are partitioned in a 2D fashion. Our B-stationary 2D SpMM uses $\mathsf{MPI\_Allgatherv}$ to replicate $\V$ tiles along each process row, avoiding $\mathsf{MPI\_Bcast}$, and uses $\mathsf{MPI\_Reduce}$ to sum partial results of $\E^T$ block rows.
This single \allgather approach is preferred over the typical $\sqrt{P}$ $\mathsf{Broadcast}$ method, as prior work \cite{sparse-summa, tripathy2020reducingcommunicationgraphneural} has shown that $\mathsf{Broadcast}$ can result in low arithmetic intensity in local SpMM computation and thus poorer performance.
In addition, because partitions of $\V$ have varying numbers of nonzeros in the 2D algorithm, the $\mathsf{Broadcast}$ approach would cause load imbalance in local SpMM.
Using \allgather avoids this imbalance by ensuring each process column contains exactly $\frac{n}{\sqrt{P}}$ nonzeros. 
This choice does not significantly affect the communication cost analysis in Section~\ref{sec:alg}, except that the $\log(\sqrt{P})$ factors in the latency and bandwidth terms are removed.
$\cv$ is computed as in the 1D routine, except the \allreduce is performed along process rows. Cluster updates use a local \argmin on each $\D^T$ partition, followed by an \mpiallreduce with $\mathsf{MPI\_MINLOC}$ along process columns. 
Finally, an \mpiallreduce along process rows computes cluster sizes, enabling a local kernel to update $\V$.

\subsection{1.5D Implementation}
\rev{
In the 1.5D implementation, $\Pp^T$ and $\Pp$ are 2D-partitioned, and $\K$ is computed using SLATE GEMM. 
$\V$ is partitioned as in the 1D algorithms.
The main difference from the 1D algorithms occurs in computing $\E^T$.
For each process row $i$, partitions of $\V$: 
\[
\begin{bmatrix} \V_{(i-1)\sqrt{P}} & \V_{(i-1)\sqrt{P}+1} & \dots & \V_{(i-1)\sqrt{P}+\sqrt{P}} \end{bmatrix},
\]
are gathered on the $i$-th diagonal process $P(i,i)$ using \mpigather. 
Each diagonal process $P(i,i)$ then uses \mpibcast to replicate the partitions of $\V$ along row $i$, and cuSPARSE SpMM computes $\E^T_{ij}(l)$.
This is equivalent to performing an $\mathsf{MPI\_Allgather}$ to replicate the partitions of $\V$ shown above along process row $i$, and has an equivalent communication cost while simplifying the implementation.
Finally, $\mathsf{MPI\_Reduce\_scatter\_block}$ is used to sum the partial results of the 1D-partitioned $\E^T$ along each process column while splitting the summed outputs along columns, yielding $\E^T$ partitioned in a 1D fashion.
Because each local SpMM produces output in row-major order, each $\E^T_{ij}(l)$ must be converted to column-major order before the $\mathsf{MPI\_Reduce\_scatter\_block}$ to ensure that the portion sent to each process is stored contiguously.
The additional time required for this conversion was negligible. 
Processes in the 2D grid are arranged in column-major order. 
This ensures that $\mathsf{MPI\_Reduce\_scatter\_block}$ along process columns naturally places the fully reduced partitions of $\E^T$ on contiguous processes, which is necessary for the 1D partitioning of $\E^T$. Computing $\D^T$ and updating cluster assignments are performed in the same way as in the 1D implementation.
}

\section{Results}\label{sec:results}

In this section, we evaluate the performance of the four \kkm algorithms presented in this paper on three real-world libSVM datasets~\cite{libsvm}, and compare the 1.5D algorithm's runtime to a single-GPU implementation of \kkm that uses a sliding-window approach to handle data that does not fit on a single GPU.
To the best of our knowledge, there are no open-source GPU-accelerated distributed \kkm implementations, so a direct comparison to a distributed baseline is not possible. However, since the 1D algorithm has a communication pattern similar to the non-linear algebraic formulations of distributed \kkm described in the literature~\cite{ferrarottietal,tsapanos2015distributed}, we use our 1D algorithm as a baseline.


\subsection{Experiment Information}

The experiments were run on the GPU partition of NERSC’s Perlmutter supercomputer~\cite{perlmutter_nersc}.
Each node has four 80 GB NVIDIA A100 GPUs\footnote{On Perlmutter, there are also nodes with 40 GB A100s, but we run on the 80 GB nodes.} connected by NVLink 3.0, with nodes interconnected via a dragonfly network and four Cassini-11 NICs.
Codes were compiled with NVCC 12.9.41 (\texttt{-O3}, C++17) and used cuSPARSE 12.9, Cray MPICH 8.1.30, and SLATE 2025.05.28.
\begin{table}[]
    \centering
    \begin{tabular}{|l|r|r|l|}
        \hline
        \rowcolor{red!20} Dataset & n\hspace{2em} & d\hspace{1.15em} & \hspace{0.35em}Domain \\
        \hline
        KDD-sampled & 8,407,752 & 10,000 & Education \\
        \rowcolor{red!10} HIGGS & 11,000,000 & 28 &  Physics \\
        MNIST8m & 8,100,000 & 784 & Vision \\
        \hline
    \end{tabular}
    \caption{The libSVM~\cite{libsvm} datasets used for evaluation.}
    \label{tab:data}
    \vspace{-1em}
\end{table}
The datasets from libSVM used to benchmark our algorithms are shown in Table \ref{tab:data}. 
The datasets were selected from different scientific domains and have varying numbers of features. 
For KDD, $10,000$ features were randomly sampled to keep $\Pp$ at a manageable size. 
This is a common practice for high-dimensional data, where dimensionality reduction or feature sampling is typically applied~\cite{dimreduction}. 
Unless otherwise noted, the experiments run the clustering loop for $100$ iterations to ensure that runtime differences arise from performance, not convergence rate.
Benchmarks use $k \in {16, 32, 64}$ and the polynomial kernel with $\gamma=1, c=1, d=2$.
Here, $G$ denotes the total number of GPUs.

\subsection{Weak Scaling}
\begin{figure}[t]
  \centering
%

  \begin{subfigure}[t]{0.45\linewidth}
    \centering
    \includegraphics[width=\linewidth]{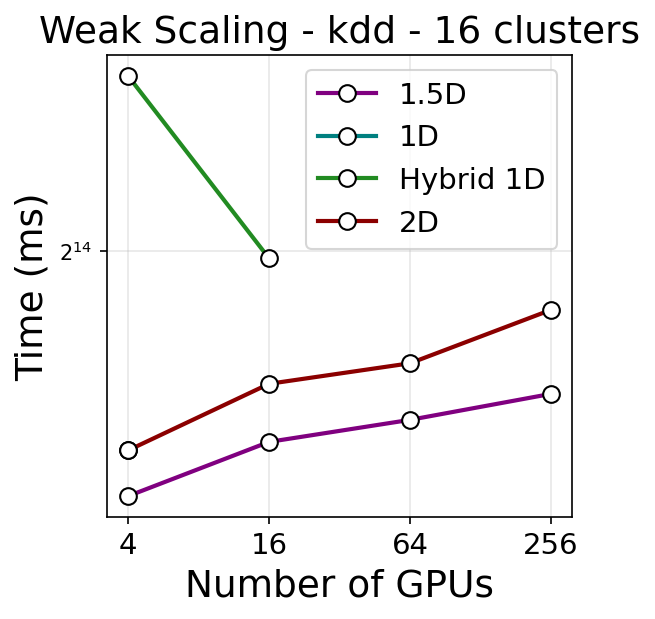}
  \end{subfigure}\hfill
  \begin{subfigure}[t]{0.45\linewidth}
    \centering
    \includegraphics[width=\linewidth]{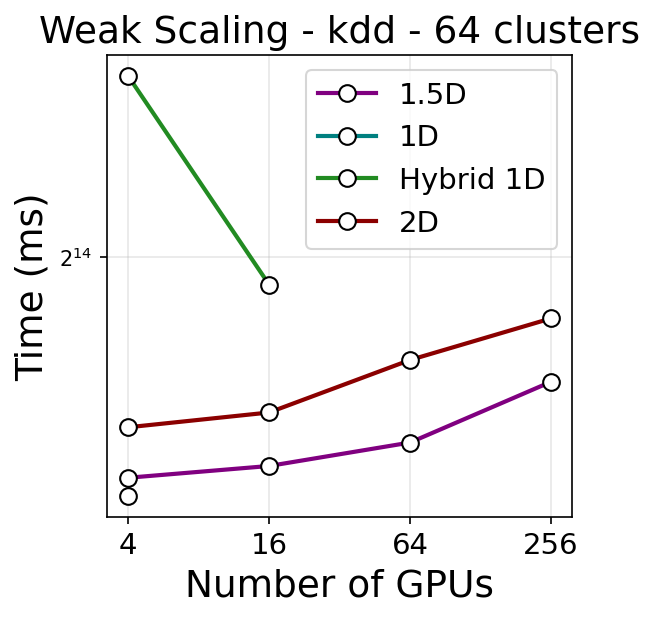}
  \end{subfigure}

  \begin{subfigure}[t]{0.48\linewidth}
    \centering
    \includegraphics[width=\linewidth]{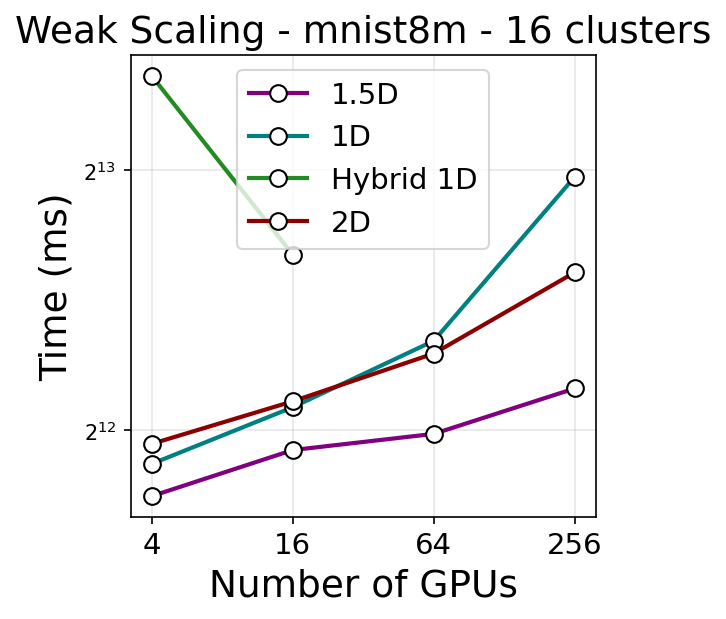}
  \end{subfigure}\hfill
  \begin{subfigure}[t]{0.48\linewidth}
    \centering
    \includegraphics[width=\linewidth]{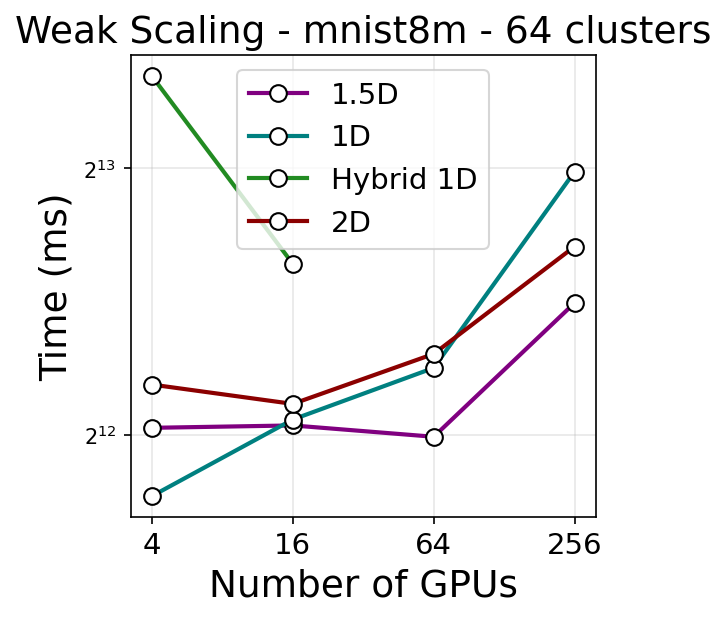}
  \end{subfigure}

  \caption{The weak scaling evaluation on three datasets and $k\in\{16,64\}$. \rev{Results for the HIGGS dataset and $k=32$ are omitted for clarity.}}
  \label{fig:weak}
\end{figure}

In this section, we evaluate the weak scaling of our \kkm algorithms. 
For each GPU count $G$, $n = \sqrt{G} \times 96{,}000$ points are sampled so that $\K$ fits within the aggregate GPU memory and the per-GPU workload for computing $\K$ and $\E^T$ remains constant as $n$ increases.
Figure~\ref{fig:weak} illustrates weak scaling for the four algorithms across datasets and $k$ values.
The 1.5D algorithm consistently achieves the best weak scaling, while the 2D algorithm outperforms both the 1D and H-1D approaches. For all $k$ values and datasets, our 1.5D algorithm achieves a geometric mean weak scaling efficiency of $86.87\%$ at 64 GPUs and $79.72\%$ at 256 GPUs. 
\rev{Our 1.5D algorithm can cluster datasets with more than 1.5 million points, representing an increase of approximately $19\times$ over the largest dataset size clustered using exact \kkm in prior work \cite{popcorn}.} 
The H-1D algorithm cannot run on more than 16 GPUs because of the extra memory needed to redistribute $\K$ from 2D to 1D. For KDD, the 1D algorithm fails on more than 4 GPUs, highlighting its memory inefficiency: the 1D GEMM routine requires replicating $\Pp$ on each GPU.
For KDD, with $d = 10{,}000$, the global $\Pp$ matrix reaches several GB as $G$ increases, making it impossible to store both a local $\K$ partition and the replicated $\Pp$ on a single GPU. In contrast, the 1.5D and 2D algorithms handle all problem sizes without memory issues, as they avoid both $\K$ redistribution and $\Pp$ replication.

\begin{figure*}
    \centering
    \begin{subfigure}[t]{0.495\textwidth}
        \centering
        \includegraphics[width=\textwidth]{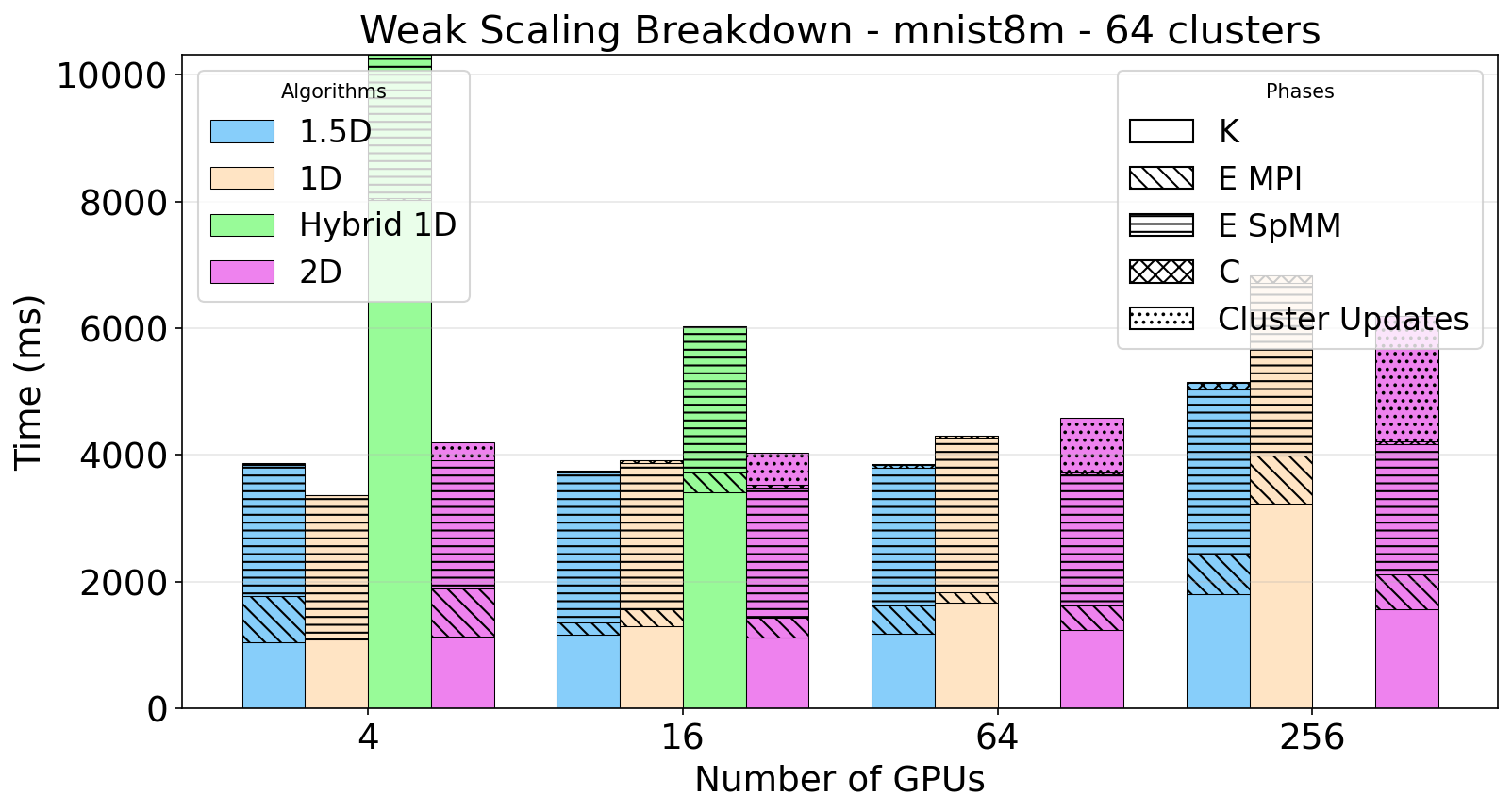}
    \end{subfigure}
    \centering
    \begin{subfigure}[t]{0.495\textwidth}
        \centering
        \includegraphics[width=\textwidth]{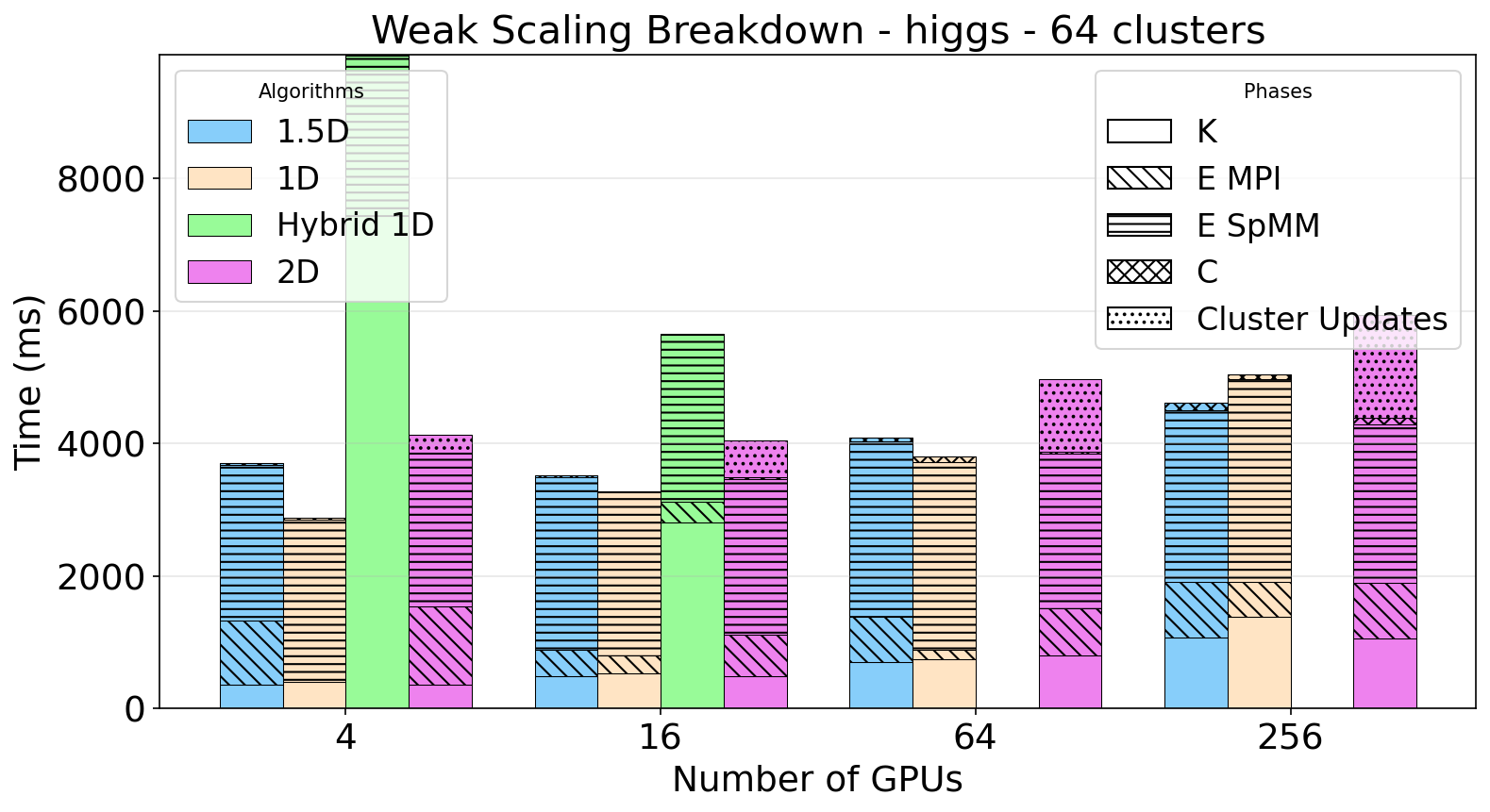}
    \end{subfigure}
    \caption{Weak scaling runtime breakdown for MNIST8m and HIGGS for $k=64$.}
    \label{fig:weak-breakdown}
    \vspace{-1em}
\end{figure*}

Figure~\ref{fig:weak-breakdown} illustrates the runtime breakdown of all four algorithms across GPU counts for MNIST8m and HIGGS with $k=64$, highlighting their performance differences. The 1D algorithm scales poorly at higher GPU counts because the $\K$ computation time increases with $G$, as discussed in Section~\ref{sec:alg}. In contrast, our 1.5D algorithm benefits from SUMMA, enabling better scalability, especially for MNIST8m, where the larger $d$ amplifies this advantage.
The communication cost of computing $\E^T$ in the 1.5D algorithm is comparable to that of the 1D algorithm for large $G$, despite the additional $\mathsf{Reduce\text{-}Scatter}$. 
This aligns with our analysis in Section~\ref{sec:alg}, which predicted that the extra cost would be negligible, especially at scale.
The H-1D algorithm is the slowest due to the high cost of computing $\K$, which is dominated by redistribution overhead.
Consequently, computing $\K$ is more expensive than in the 1D case, even though a more communication-efficient algorithm is used.

The 2D algorithm scales similarly to the 1.5D algorithm but generally performs worse due to additional communication during cluster updates, particularly the $\mathsf{MPI\_Allreduce}$ with $\mathsf{MPI\_MINLOC}$. 
For large $G$, this overhead becomes significant. 
Our analysis showed that $\mathsf{MPI\_Allreduce}$ communicates $\cO(\log(\sqrt{P})n)$ words. 
The $\mathsf{MPI\_MINLOC}$ operator also doubles the buffer size to store an additional integer, further increasing the communication volume.

\subsection{Strong Scaling}

\begin{figure}[t]
  \centering
%

  \begin{subfigure}[t]{0.45\linewidth}
    \centering
    \includegraphics[width=\linewidth]{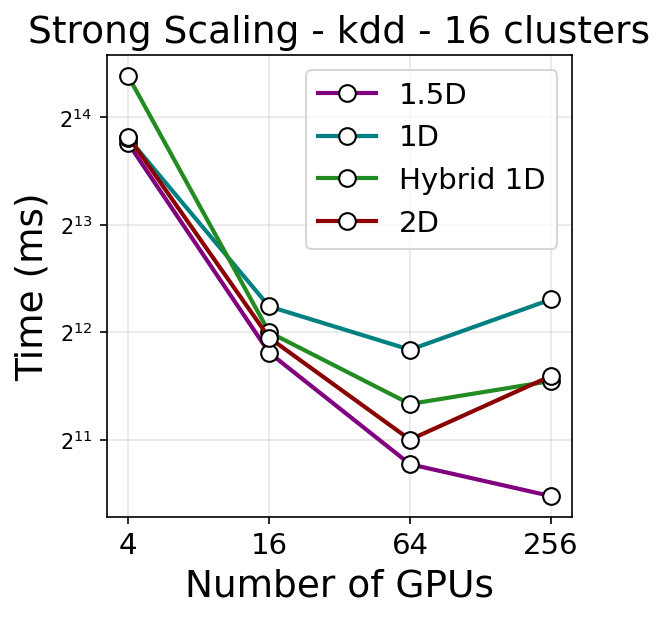}
  \end{subfigure}\hfill
  \begin{subfigure}[t]{0.45\linewidth}
    \centering
    \includegraphics[width=\linewidth]{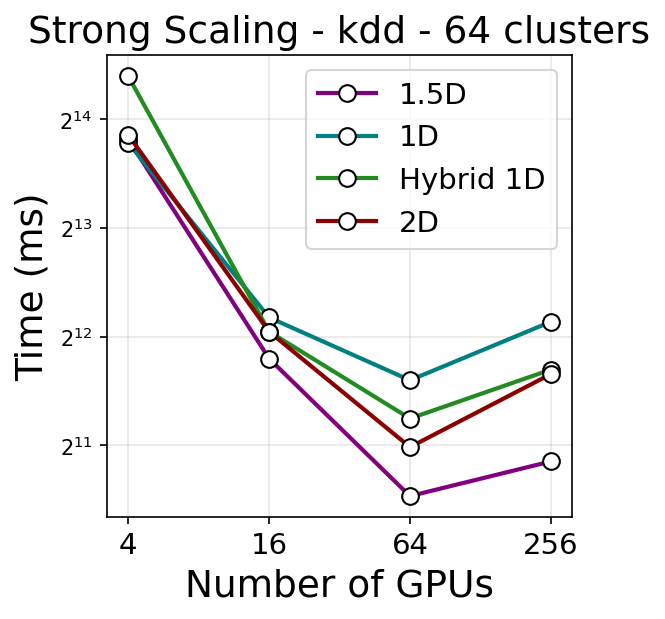}
  \end{subfigure}

  \vspace{0.6em}

  \begin{subfigure}[t]{0.48\linewidth}
    \centering
    \includegraphics[width=\linewidth]{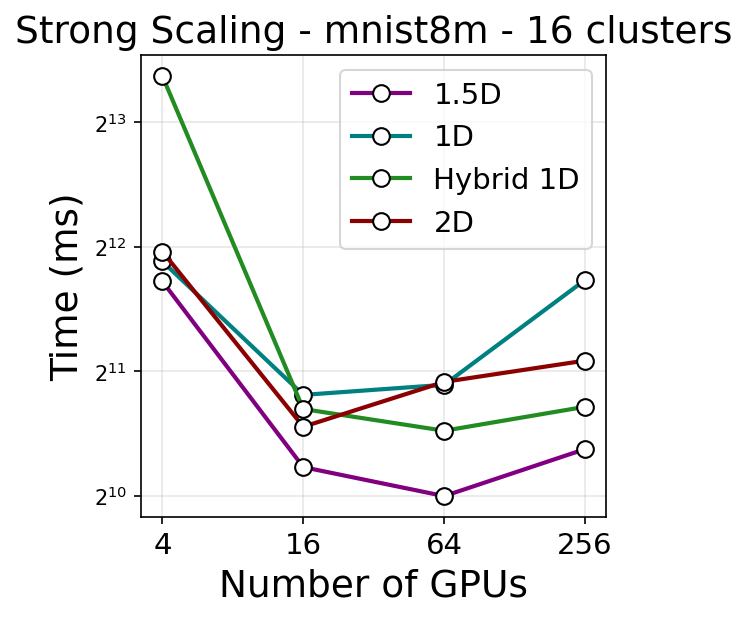}
  \end{subfigure}\hfill
  \begin{subfigure}[t]{0.48\linewidth}
    \centering
    \includegraphics[width=\linewidth]{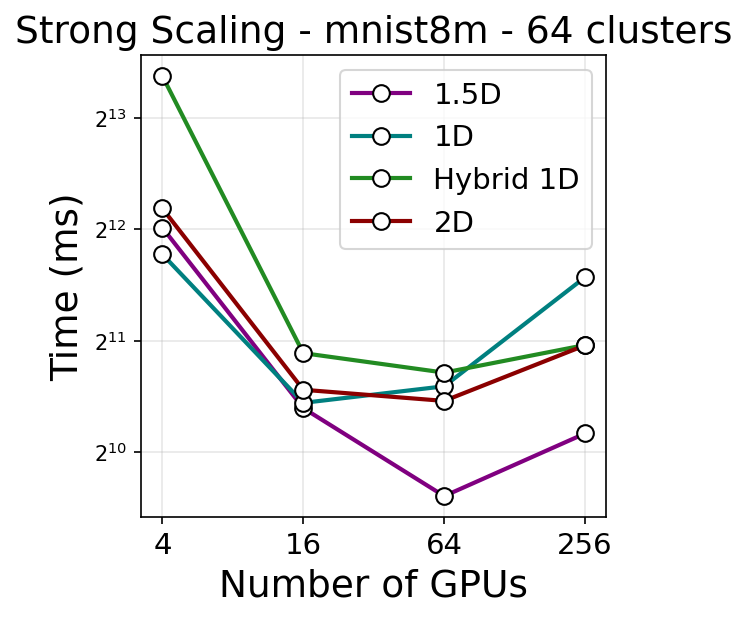}
  \end{subfigure}

  \caption{Strong scaling evaluated on three datasets with $k \in \{16, 64\}$. \rev{Results for the HIGGS dataset and $k=32$ are omitted for clarity.}}
  \label{fig:strong-scaling}
  \vspace{-1em}
\end{figure}

This section evaluates the strong scaling of our \kkm algorithms using $n=192{,}000$ sampled points per dataset, resulting in a kernel matrix near the single-node memory limit.
Figure~\ref{fig:strong-scaling} illustrates strong scaling results for the four algorithms across datasets and $k$ values. 
The 1.5D algorithm consistently scales best. Its advantage is most evident at lower $k$ values (e.g., $k=16$), where it scales efficiently to 256 GPUs on KDD.
For MNIST8m and larger $k$ values, the 1.5D algorithm generally does not scale beyond 64 GPUs but still outperforms others in scalability and runtime. Overall, across all values of $k$ and all datasets, the 1.5D algorithm achieves a geometric mean strong scaling speedup of $4.65\times$ at 64 GPUs and $4.16\times$ at 256 GPUs. Both the 2D and H-1D algorithms consistently scale better than the 1D algorithm.

\begin{figure*}
    \begin{subfigure}[t]{0.495\textwidth}
        \centering
        \includegraphics[width=\textwidth]{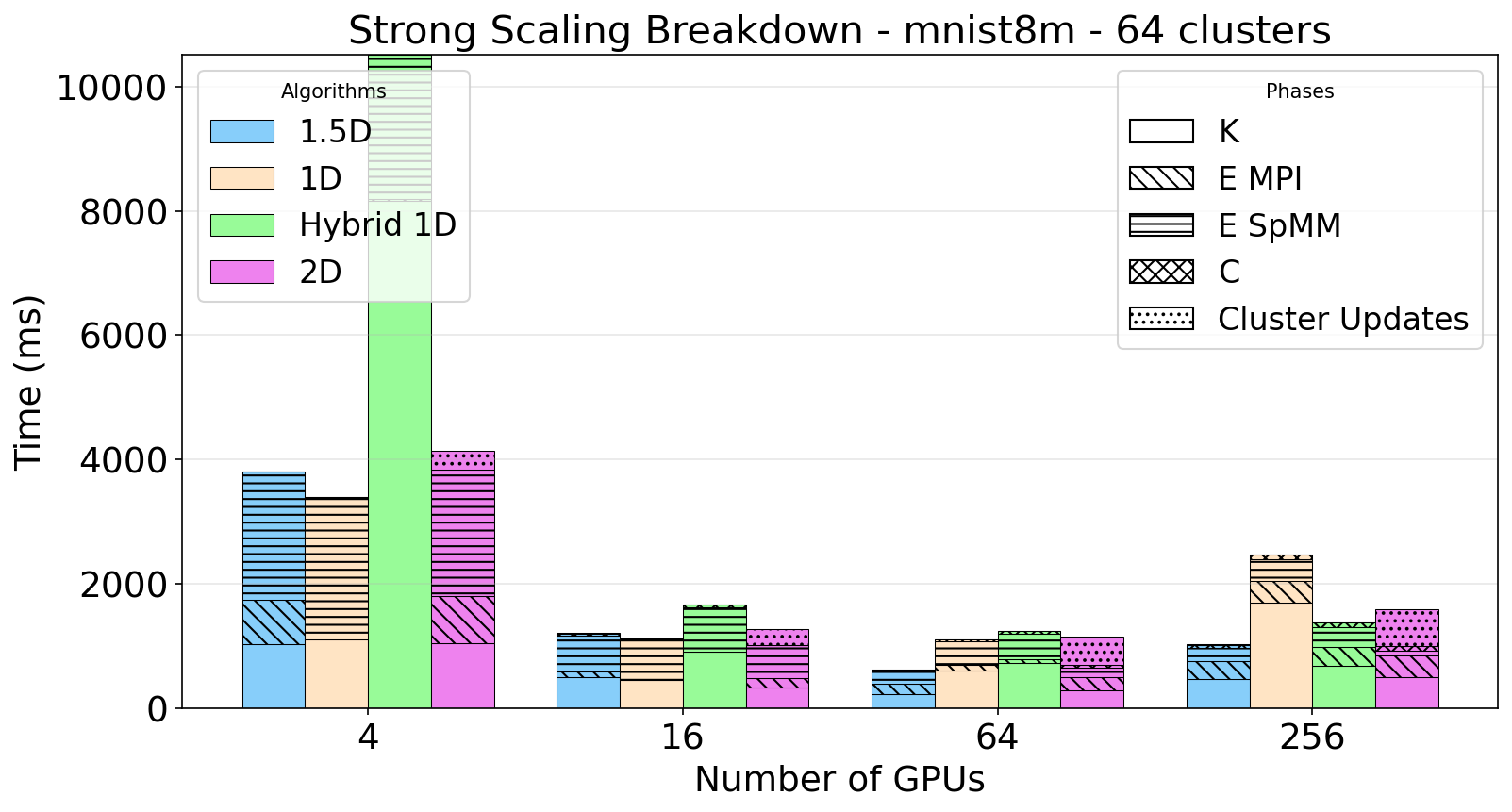}
    \end{subfigure}
    \centering
    \begin{subfigure}[t]{0.495\textwidth}
        \centering
        \includegraphics[width=\textwidth]{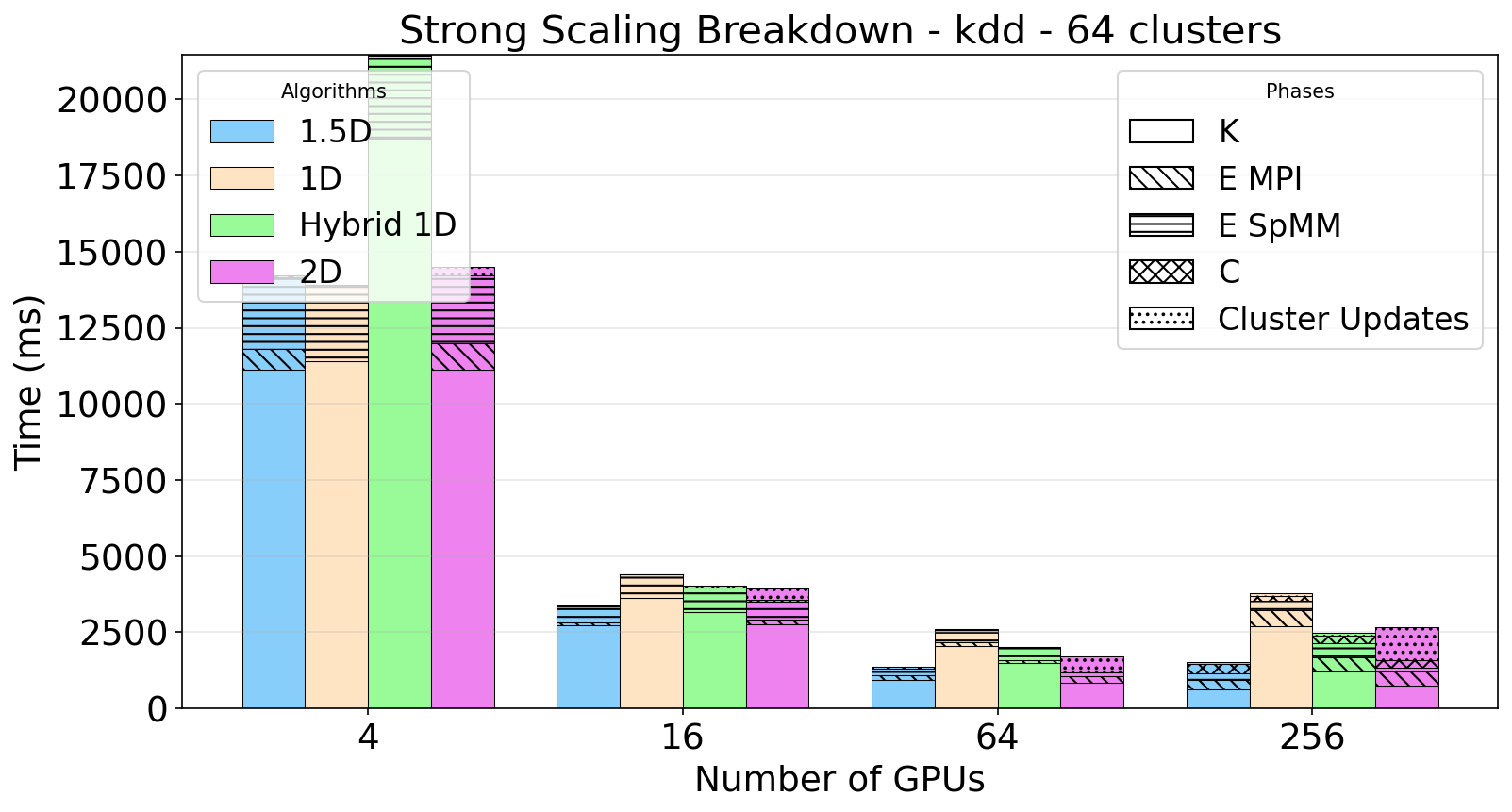}
    \end{subfigure}
    \caption{The strong scaling runtime breakdown on MNIST8m and KDD for $k=64$.}
    \label{fig:strong-breakdown}
    \vspace{-1em}
\end{figure*}

Figure~\ref{fig:strong-breakdown} illustrates the runtime breakdown for strong scaling on MNIST8m and KDD with $k=64$. 
Other datasets and other values of $k$ show similar patterns to the ones in these plots. 
As in the weak scaling case, the 1D algorithm is limited by poor $\K$ scalability, while the 1.5D algorithm can avoid that bottleneck. 
The additional communication required to compute $\E^T$ in 1.5D is minimal and scales well, eventually matching the 1D algorithm's communication cost.
In strong scaling, the $\cO(\frac{n^2}{\sqrt{P}})$ redistribution cost in H-1D scales with GPU count, but the $\cO(P)$ message count still creates a latency bottleneck, keeping it behind 1.5D. 
The 2D algorithm faces the same issue as in weak scaling: the $\mathsf{MPI\_Allreduce}$ for \argmin does not scale with GPU count and eventually becomes a bottleneck.


\subsection{Comparison to Single GPU}

Finally, we compare our 1.5D algorithm to a single-GPU sliding-window baseline for cases where $\K$ exceeds GPU memory, using $n = 192{,}000$. 
The 1.5D approach is emphasized because it consistently outperforms other strategies.

Our sliding window algorithm is based on the approach to \kkm described in \cite{zhang2002large}, which stores $\K$ on disk and loads it in blocks. In contrast, we recompute blocks of $\K$ on the fly, trading increased computation for reduced disk I/O and host-device data movement, both of which are more costly than computation.
The sliding window algorithm computes a $b \times n$ block row of $\K$ and updates $b$ rows of $\D$ at each step. 
Once $\lceil n/b \rceil$ steps are completed, new cluster assignments ($cl_p$ and $\V$) are computed before the next \kkm iteration. 
We set $b = 8192$, which yielded the best empirical performance.
Computing $\K_p$ is the main bottleneck in the sliding window algorithm because of the cost of GEMM and potential nonlinear operations, such as exponentiation for the polynomial kernel. 
Therefore, we store $\V$ as a dense matrix in our experiments, since the time required to recompute $\K_p$ dominates the computation of $\E^T$.
Popcorn~\cite{popcorn} was considered as a single-GPU baseline, but it failed to cluster datasets larger than approximately $80,000$ points, making it unsuitable for our large-scale focus. 
To the best of our knowledge, no open-source distributed \kkm implementations with GPU support exist. 
Therefore, the sliding window algorithm serves as the most relevant baseline outside our 1D algorithm.

\begin{figure*}[t]
    \centering
    \begin{subfigure}[b]{0.32\textwidth}
        \centering
        \includegraphics[width=\textwidth]{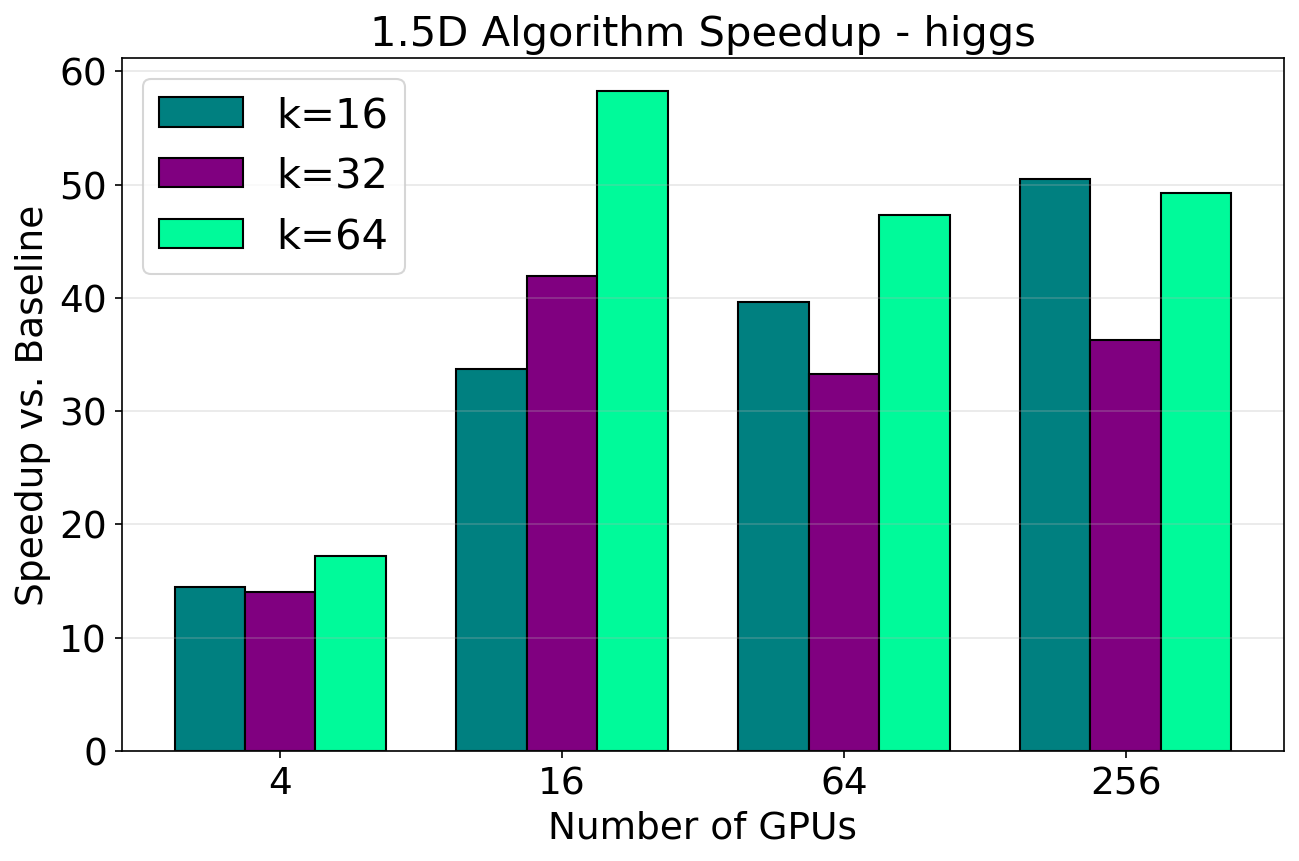}
    \end{subfigure}
    \hfill
    \begin{subfigure}[b]{0.32\textwidth}
        \centering
        \includegraphics[width=\textwidth]{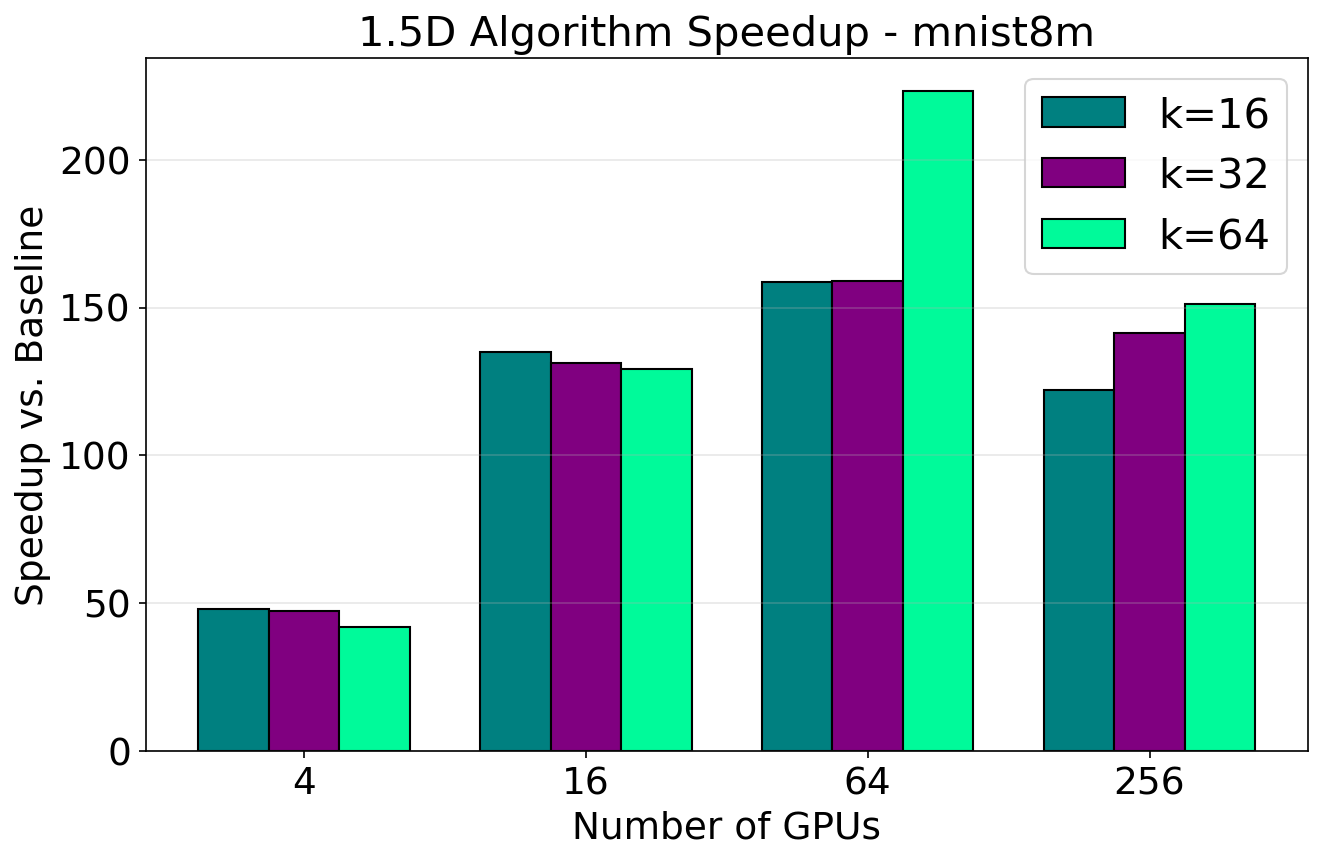}
    \end{subfigure}
    \hfill
    \begin{subfigure}[b]{0.32\textwidth}
        \centering
        \includegraphics[width=\textwidth]{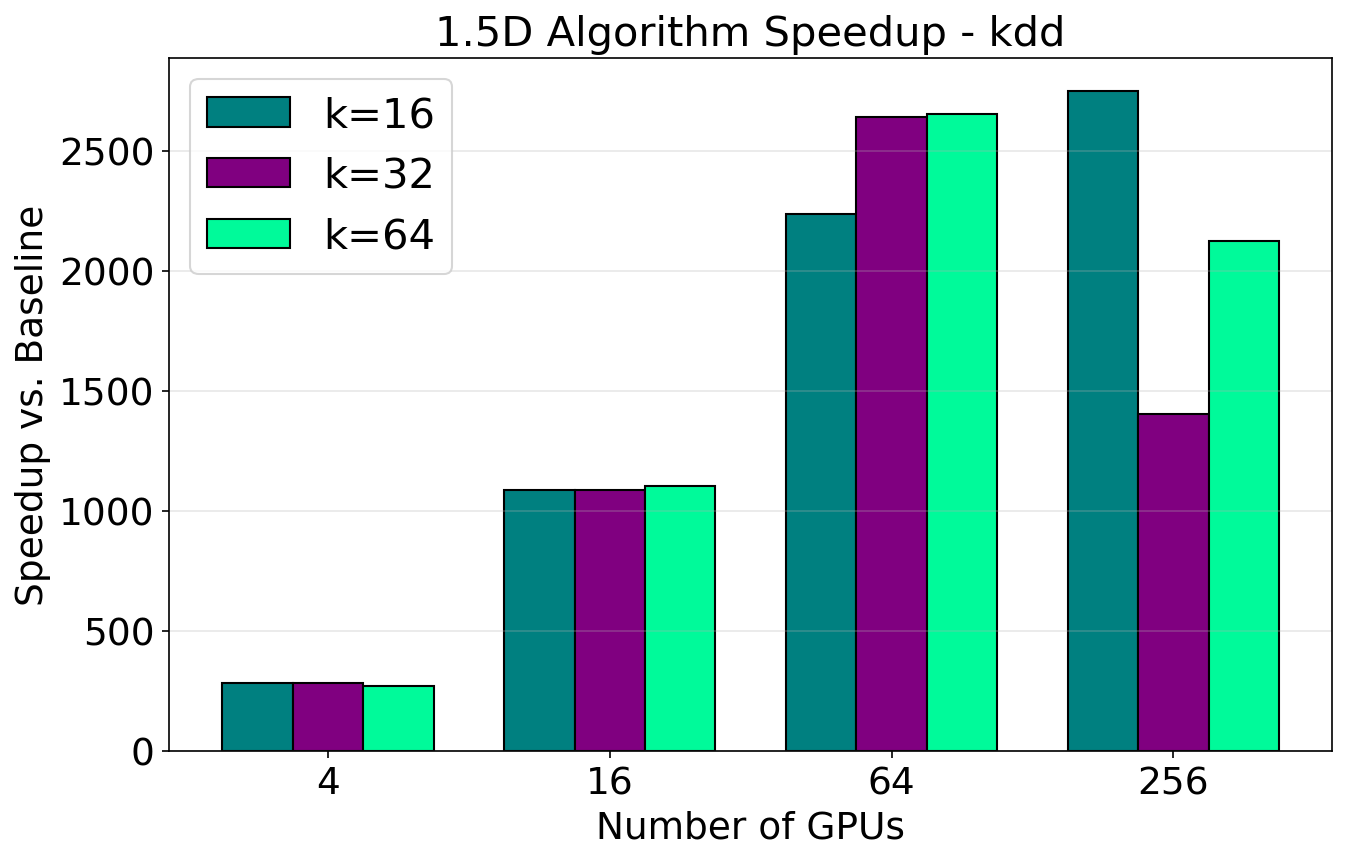}
    \end{subfigure}
    \caption{The evaluation of strong scaling speedup over sliding window approach.}
    \label{fig:speedup}
    \vspace{-1em}
\end{figure*}

Figure~\ref{fig:speedup} illustrates the speedup of the 1.5D algorithm over the sliding window baseline across datasets and $k$ values. 
At 256 GPUs, 1.5D is more than $10\times$ faster in all cases, with the largest gain on KDD at $k=16$, achieving a $2749.8\times$ speedup and reducing runtime from over an hour to under 2 seconds. 
The speedups are especially pronounced for datasets with large $d$, where recomputing $\K_p$ in the sliding window approach is more costly.
Being able to cluster large datasets with \kkm in seconds instead of hours significantly expands the scope of potential applications of \kkm, since large-scale clustering in reasonable time is now possible.

\section{Conclusion and Future Work}

This work presents GPU-accelerated distributed-memory algorithms for \kkm based on sparse and dense linear algebra primitives. 
By tailoring communication strategies and partitioning schemes to the structure of this clustering algorithm, particularly the sparsity of $\V$ and the interaction between GEMM and SpMM, we make large-scale exact \kkm clustering practical. 
In contrast to previous single-GPU exact approaches, our method enables clustering of datasets that are one to two orders of magnitude larger within reasonable time.

In this work, we study four distributed formulations: a 1D algorithm, a hybrid 1D algorithm, a 2D algorithm, and a 1.5D algorithm. 
The proposed 1.5D algorithm minimizes communication volume and achieves better scalability, with communication decreasing as the number of processes increases, unlike existing approaches where communication remains constant or increases more rapidly. 
\rev{Its effectiveness comes from the application-aware composability of linear algebra primitives: 1D-partitioned $\V$ enables SpMM to consume the 2D-partitioned $\K$ from SUMMA without redistribution, and 1D-partitioned $\mathbf{E}^T$ eliminates communication for cluster updates.}
As a result, our 1.5D algorithm demonstrates strong and weak scaling up to 256 GPUs on large real-world datasets, achieving substantial speedups over a single-GPU sliding-window baseline. 
Most importantly, it enables exact \kkm clustering of datasets with more than 1.5 million points, representing a dramatic increase in tractable problem size compared to prior work.


For future work, we plan to develop additional clustering algorithms, such as spectral clustering and mean-shift, using sparse linear algebra, and implement them on distributed memory systems.
In this scenario, we plan to explore mixed-precision techniques for \kkm, standard K-means, and related clustering algorithms to ensure portability to newer architectures favoring lower precision.
\rev{More broadly, we aim to develop methodologies that leverage domain-specific sparsity structures and sequences of linear algebra primitives to select communication-efficient data distribution and implementation strategies. 
One possible direction is an autotuning framework that analyzes end-to-end sequences of sparse and dense linear algebra primitives to automatically select optimal data distributions and communication schedules for the entire workflow.
}

\section*{Acknowledgment}

This research used resources of the National Energy Research Scientific Computing Center, a DOE Office of Science User Facility supported by the Office of Science of the U.S. Department of Energy under Contract No. DE-AC02-05CH11231, using NERSC award ASCR-ERCAP0030076. 
This material is based upon work supported by the U.S. Department of Energy, Office of Science, Office of Advanced Scientific Computing Research, Department of Energy Computational Science Graduate Fellowship under Award Number DE-SC0025528.
The authors acknowledge financial support from \textit{ICSC – Centro Nazionale di Ricerca in High-Performance Computing, Big Data and Quantum Computing}, funded by the European Union -- NextGenerationEU.
The authors wish to disclose that generative AI and editing assistants, such as Microsoft Copilot and InstaText, were used to assist with grammar checking and to improve the clarity of the writing.

\bibliographystyle{plain}
\bibliography{bibliography}

\end{document}